\documentclass[aps,prd,twocolumn,superscriptaddress,showpacs,showkeys,nofootinbib]{revtex4-1} 

\usepackage{latexsym}
\usepackage{amsmath}
\usepackage{amssymb}
\usepackage{amsfonts}
\usepackage{slashed}
\usepackage{bbm}
\usepackage{color}

\usepackage{supertabular} 
\usepackage{placeins}
\usepackage{epsfig}
\usepackage{graphicx}
\usepackage{multirow}

\newcommand{\mbare}{\stackrel{\circ}{m}_{c\bar s}}
\newcommand{\comp}{\mathbbm{C}}
\newcommand{\reales}{\mathbbm{R}}
\newcommand{\Dw}{D_s^{(\ast)}}
\newcommand{\Dx}{D^{(\ast)}}

\newcommand{\Dvd}{D^\ast_{s0}(2317)}
\newcommand{\Dsu}{D_{s1}(2460)}

\usepackage[dvipsnames]{xcolor}
\usepackage[colorlinks=true,allcolors=BlueViolet]{hyperref}

\begin{document}

\frenchspacing


\title{\boldmath Contribution of constituent quark model $c\bar s$ states to the dynamics of the $D_{s0}^\ast(2317)$ and $D_{s1}(2460)$ resonances}



\author{Miguel Albaladejo}
\email[]{albaladejo@um.es}
\affiliation{Departamento de F\'isica, Universidad de Murcia, E-30071 Murcia, Spain}

\author{Pedro Fernandez-Soler}
\email[]{pedro.fernandez@ific.uv.es}

\author{Juan Nieves}
\email[]{jmnieves@ific.uv.es}
\affiliation{Instituto de F\'isica Corpuscular (IFIC), 
CSIC-Universidad de Valencia, E-46071 Valencia, Spain}

\author{Pablo G. Ortega}
\email[]{pgortega@usal.es}
\affiliation{Grupo de F\'isica Nuclear, 
Universidad de Salamanca, E-37008 Salamanca, Spain}


\date{\today}

\begin{abstract}
The masses of the $\Dvd$ and $\Dsu$ resonances lie below the $DK$ and $D^\ast K$ thresholds respectively,  which contradicts the 
predictions of naive quark models and points out to 
non-negligible effects of the $D^{(\ast)}K$ loops in the dynamics of the even-parity scalar ($J^\pi=0^+$) and axial-vector ($J^\pi=1^+$) 
$c\bar s$ systems.  Recent lattice QCD studies, incorporating the 
effects of the $D^{(\ast)}K$ channels, analyzed these spin-parity sectors and correctly described  the $D^\ast_{s0}(2317)-D_{s1}(2460)$ mass splitting.
Motivated by such works, we study the structure of the $D_{s0}^\ast(2317)$ and $D_{s1}(2460)$
resonances in the framework of an effective field theory consistent with 
heavy quark spin symmetry, and that incorporates the interplay between $D^{(\ast)}K$ meson-meson degrees of freedom and bare P-wave $c\bar s$ states predicted by constituent quark models. We extend the scheme to finite volumes and fit the strength of the coupling between both types of degrees of freedom to the available lattice levels, which we successfully describe. We finally estimate the size of the $D^{(\ast)}K$ two-meson components in the $D^\ast_{s0}(2317)$ and $D_{s1}(2460)$ resonances, and we conclude that these states have a predominantly hadronic-molecular structure, and that it should not be tried to accommodate these mesons within $c\bar{s}$ constituent quark model patterns.
\end{abstract}



\maketitle



\section{Introduction}
\label{sec:introduction}

The $D_{s0}^{\ast}(2317)$ and $D_{s1}(2460)$ mesons were discovered in $2003$, the first one by the \textsc{BaBar} Collaboration 
in the $D_s^+\pi^0$ invariant mass spectrum with $J^\pi=0^+$ quantum numbers~\cite{Aubert:2003fg} and the second one by the \textsc{Cleo} Collaboration
in the $J^\pi=1^+$ sector analyzing the $D_s^{\ast\,+}\pi^0$ channel~\cite{Besson:2003cp}.
The states quickly raised attention due to their narrow widths and low masses: 
the $D_{s0}^\ast(2317)$ meson has a width of $\Gamma<3.8$ MeV ($95\%$ CL) and a mass of $2317.7\pm 0.6$ MeV, $40$ MeV below the $DK$ threshold, while the
$D_{s1}(2460)$ width and mass are $\Gamma<3.5$ MeV ($95\%$ CL) and $M=2459.5\pm 0.6$ MeV, $45$ MeV below the $D^\ast K$ threshold. Such low values could not be accommodated in the 
predictions of the so far fruitful quark models~\cite{Godfrey:1986wj, 
Zeng:1994vj, Gupta:1994mw, Ebert:1997nk, Lahde:1999ih, DiPierro:2001dwf} and lattice QCD 
calculations~\cite{Boyle:1997aq, Boyle:1997rk, Hein:2000qu, Lewis:2000sv, Bali:2003jv, 
diPierro:2003iw, Dougall:2003hv}, that expected these resonances to lie well above the respective $D^{(\ast)}K$ thresholds.

The presence of the heavy charm quark in the $D_s$ states implies the validity, up to 
$\Lambda_\text{QCD}/m_Q$ corrections, of Heavy Quark Spin Symmetry 
(HQSS)~\cite{Isgur:1989vq, Isgur:1989ed, Georgi:1990um, Isgur:1991wq,Manohar:2000dt}, with $m_Q$ the heavy quark mass (the charm quark mass in this case),  and $\Lambda_\text{QCD}$ a typical scale related to the dynamics of the light degrees of freedom. Thus, in good approximation, the spin 
of the heavy quark $s_Q$ is decoupled from the 
total angular momentum of the light degrees of freedom $j_{\bar q}$, and hence they are 
separately conserved. This gives rise to the arrangement of charmed-strange mesons in doublets, classified by 
the total angular momentum and parity,  $j^\pi_{\bar q}$, of their light degrees of freedom content, and with total spin $J=j_{\bar q}\pm 1/2$ and parity $\pi$. For the P-wave $D_s$ mesons, the expected HQSS 
doublets are,\footnote{The parity of the light degrees of freedom in this case is $+$, that corresponds to an odd parity $\bar s$ antiquark orbiting in P wave around the heavy  quark $c$, while the total angular momentum of the light degrees of freedom is determined by the sum of the spin 1/2 of the $\bar s$ antiquark and its orbital  angular  momentum ($\ell =1$).} on the one hand, $j_{\bar q}^\pi=\frac{1}{2}^+$ with $J^\pi=0^+,1^+$ mesons, which in S-wave couple to $DK$ and $D^\ast K$, respectively;
and, on the other 
hand, $j_{\bar q}^+=\frac{3}{2}^+$ with $J^\pi=1^+,2^+$ mesons, where the $1^+$ ($2^+$) meson can couple to $D^\ast K$ ($DK$ and  $D^\ast K$) in D-wave. Axial $1^+$ states will also couple to $D^*K$ states in D-wave. In addition,  
$D^{(\ast)}K^\ast$ pairs   in S and D waves might also couple to the latter $D_s$ states, but however the dynamics of these pairs will not be governed by chiral symmetry since light-vector mesons are involved.

Experimentally, the positive parity $j_{\bar q}=\tfrac{1}{2}$ doublet would be composed of the $D_{s0}^{\ast}(2317)$ and $D_{s1}(2460)$ mesons, which were expected to be almost degenerated and broad, decaying to $D^{(\ast)}K$ through S-wave. However, neither of these two properties are empirically observed.
This caused an intense debate on the nature of the resonances, producing a wide variety of 
interpretations among which we highlight their assignment to  $c\bar s$ 
states~\cite{Fayyazuddin:2003aa, Sadzikowski:2003jy, Lakhina:2006fy} and 
two-meson or four-quark systems~\cite{Barnes:2003dj, 
Lipkin:2003zk, Szczepaniak:2003vy, Kolomeitsev:2003ac,
  Hofmann:2003je,  Browder:2003fk, Bicudo:2004dx, Bicudo:2004dx,Guo:2006fu, Gamermann:2006nm,Faessler:2007gv,Flynn:2007ki, MartinezTorres:2011pr, Altenbuchinger:2013vwa, Altenbuchinger:2013gaa}.

The latest lattice QCD (LQCD) simulations~\cite{Bali:2017pdv,Mohler:2013rwa,Lang:2014yfa} have achieved a good description of these charmed-strange resonances when $DK$ interpolators are included in the set of used operators.  Notably, the mass of the $D_{s0}^\ast(2317)$ was found to be overestimated if the $DK$ interpolators were omitted, which gives 
further support to the idea of a necessary interplay between constituent quark model (CQM) configurations and nearby $D^{(\ast)}K$ thresholds.

On the other hand, if one accepts the predictions of generally successful CQMs, one should expect the charmed-strange  $J^\pi=0^+$ ground state to lie much closer to the $DK$ threshold than the physical $D_{s0}^{\ast}(2317)$, so the latter meson pair could act as an essential dynamical agent to reduce the mass of the bare meson state closer to the experimental value, as suggested by some authors~\cite{vanBeveren:2003kd}. Hence, in this picture, the physical $D_{s0}^\ast(2317)$ resonance would be the result of a strong renormalization of a bare $c\bar s$ component, rather than a new dynamical state generated from a strongly attractive $DK$ interaction. Nevertheless, since the required renormalization would be quite significant, one would expect, even in this context, that the $D_{s0}^\ast(2317)$ resonance will acquire a sizable two-meson molecular probability. Indeed, the low-lying P-wave charmed-strange mesons were  studied in  Ref.~\cite{Ortega:2016mms} employing a widely used CQM~\cite{Vijande:2004he,
Valcarce:2005em,Segovia:2013wma}, where the coupling between the quark-antiquark and meson-meson degrees of freedom was modeled with the $^3P_0$ transition operator~\cite{LeYaouanc:1972ae}. In that work, where all the parameters were constrained from previous studies on hadron phenomenology, the bare $1^3P_0$ $c\bar s$ state\footnote{We use the nomenclature $n\, ^{2S+1}L_{2J+1}$ for the radial, spin, orbital and total angular momentum quantum numbers of a quark-antiquark state. In addition the parity of the state is $(-1)^{L+1}$.} developed a large mass-shift as a consequence of its coupling with the $DK-$meson pair, becoming its mass closer to that of the physical $D_{s0}^\ast(2317)$ resonance. On the other hand, the dressed state contained a large molecular component that gave rise to a $DK-$meson pair probability of  around $33\%$~\cite{Ortega:2016mms} in the final configuration of the meson.

In sharp contrast, the LQCD energy-levels reported in Refs.~\cite{Mohler:2013rwa, Lang:2014yfa} were analyzed in Ref.~\cite{Torres:2014vna}, employing  an auxiliary potential method, where  $DK$ molecular probabilities for the $D_{s0}^{\ast}(2317)$ much higher,  of the order of $70\%$, were found. This result was consistent with the previous values obtained in  Ref.~\cite{Liu:2012zya}. The authors of this  work  performed a LQCD calculation of several heavy-light meson--Golstone boson scattering lengths, that they used to fit the  LECs entering in the unitarized next-to-leading (NLO) heavy meson chiral perturbation theory  (HMChPT) coupled-channel $T-$matrix derived in Ref.~\cite{Guo:2008gp}. The latter amplitudes were employed to estimate the $\Dvd$ molecular component. These high values for the $DK$ probabilities\footnote{Note that Ref.~\cite{Liu:2012zya} made use of the Weinberg compositeness rule~\cite{Weinberg:1965zz}, which relates this probability to the scattering length in the limit of small binding 
energies. However, the works of Ref.~\cite{Torres:2014vna,Albaladejo:2016hae} employed a generalization~\cite{Gamermann:2009uq} that remains valid for bound states, independently of their distance to threshold.}  are similar to those obtained in Ref.~\cite{Albaladejo:2016hae} from the analysis of the experimental $DK$ invariant mass spectra of the reactions $B^+ \to \bar D^0 D^0 K^+$, $B^0 \to D^-D^0 K^+$ \cite{Lees:2014abp}  and $B_s \to \bar D^0 K^-\pi^+$ \cite{Aaij:2014baa} measured by the BaBar and LHCb Collaborations, respectively. In all cases an enhancement right above the threshold is seen and it is related in Ref.~\cite{Albaladejo:2016hae} to the presence of the $D^*_{s0}(2317)$. The latter is dynamically generated  when the leading order (LO) HMChPT amplitudes are used as the kernel of a Bethe-Salpeter equation (BSE), whose renormalized solutions fulfill exact elastic unitarity in coupled-channels. 

The predominantly molecular structure of the $D_{s0}^{\ast}(2317)$ and $D_{s1}(2460)$ resonances has recently received  an indirect robust 
theoretical support in Refs.~\cite{Albaladejo:2016lbb, Du:2017zvv} (see also the review of Ref.~\cite{Guo:2017jvc}). In the first of these two references,
the heavy-light pseudoscalar meson $J^\pi=0^+$ scattering in the strangeness-isospin $(S,I)=(0,1/2)$ sector was studied, and a
strong case for the existence of two poles in the $D_0^*(2400)$ energy-region was presented. The affirmative
evidence came from a remarkably good agreement between a {\em parameter-free predicted} low-lying levels calculated using NLO HMChPT unitarized amplitudes~\cite{Guo:2008gp} and the LQCD results
reported in Ref.~\cite{Moir:2016srx}. The dynamical origin of this two-pole structure was 
elucidated from the light-flavor SU(3) structure of the interaction, and it was found that the lower pole would be the SU(3) partner of the
$D^\ast_{s0}(2317)$. Thus, this latter state will have also clear hadron-molecular origin. A similar pattern was found for $J^\pi=1^+$ and in the bottom sector. This in fact might solve a  long-standing puzzle in
charm-meson spectroscopy, since it would provide an explanation of why the  masses, quoted in the PDG~\cite{Olive:2016xmw},  of the non-strange mesons $D^*_0(2400)$ and $D_1(2430)$ are almost equal to or even
higher than their strange siblings. In  the second reference~\cite{Du:2017zvv},  it is shown that the well-constrained amplitudes for Goldstone bosons scattering off charm mesons used in Ref.~\cite{Albaladejo:2016lbb}
are fully consistent with recent high quality data on the $B^- \to D^+\pi^-\pi^-$ and $B_s^0 \rightarrow \bar{D}^0 K^-\pi^+$ final states provided by the LHCb experiment in Refs.~\cite{Aaij:2016fma} and \cite{Aaij:2014baa}, respectively.  Indeed, all these results 
suggest a new paradigm for heavy-light meson spectroscopy that questions their traditional $q\bar q$  CQM interpretation~\cite{Du:2017zvv}.

Most of these latter works~\cite{Liu:2012zya, Albaladejo:2016hae, Albaladejo:2016lbb, Du:2017zvv} do not incorporate explicitly the bare $c\bar s$ degrees of freedom, whose effects are, in principle, encoded in the low energy constants (LECs) that appear beyond  LO in the chiral expansion, and in the non-perturbative re-summation employed to restore elastic coupled-channels unitarity. However, and depending on the proximity of the CQM states to the energy-region under study, this approximation might not be sufficiently accurate.

Such radically different pictures of the inner structure of the $D_{s0}^\ast(2317)$ makes timely a re-analysis of this resonance, paying special attention to the interplay between meson molecular and CQM degrees of freedom. We will employ here the scheme  designed in Ref.~\cite{Albaladejo:2016ztm} for its bottom heavy-flavor partner, and we will try to describe the charm $0^+$ and $1^+$ LQCD energy-levels obtained in Ref.~\cite{Bali:2017pdv}. Such comparison will also serve to further\footnote{As it will be discussed below, LECs in the charm and bottom sectors might be related by heavy-quark flavor symmetry.} constrain/determine the LECs that appear in the approach. The scheme of Ref.~\cite{Albaladejo:2016ztm} started from a unitary ansazt for the heavy-light-meson-Goldstone-boson $0^+$ and $1^+$ amplitudes, based on LO HMChPT $\bar B^{(*)} K $ interactions, and computed for finite volumes.  In addition,  and for the very first time in this context, the two-meson channels were coupled to the corresponding 
CQM P-wave $\bar B_s$ scalar and axial mesons using an effective interaction consistent with HQSS. In the $m_Q\to\infty$ limit and besides HQSS, Quantum Chromodynamics (QCD) acquires  also an approximate heavy flavour symmetry  that ensures that the dynamics of the system containing a single heavy quark is, up to $\mathcal{O}(\Lambda_{\rm QCD}/m_Q)$ corrections, independent of the flavor of the heavy quark~\cite{Manohar:2000dt}. As a consequence, the bottom-strange and charm-strange systems are expected to have a similar behavior. Hence, 
the  analysis of the $B_s$ low-lying spectrum carried out in Ref.~\cite{Albaladejo:2016ztm}, can be readily used here to study the charmed-strange $j_{\bar q}^\pi=\tfrac{1}{2}^+$ HQSS doublet.

The recent LQCD simulation of Ref.~\cite{Bali:2017pdv} reported finite volume energy-levels from a high statistics study of the $J^\pi=0^+$ and $1^+$ charmed-strange mesons, $D^*_{s0}(2317)$ 
and $D_{s1}(2460)$, respectively, where  the effects of the nearby $DK$ and $D^\ast K$ thresholds were taken into account by employing the corresponding four-quark operators. As we will discuss below, the work of Ref.~\cite{Bali:2017pdv}   represents a clear improvement on the pioneering ones of Refs.~\cite{Mohler:2013rwa, Lang:2014yfa}. Some of the energy-levels reported in Ref.~\cite{Bali:2017pdv} lie in the proximity of, when not above, the expected CQM  bare masses of the ground scalar and axial charmed-strange states, being thus interesting  to include explicitly these degrees of freedom in the scheme, since their effects might not be properly taken into account by simply including LECs.  Moreover, in the present study, we will also include the next $(S=1,I=0)$ higher thresholds, $D_s^{(*)}\eta$,  since they appear at energies below some of the finite-volume levels computed in Ref.~\cite{Bali:2017pdv}. 

This work is structured  as follows. After this Introduction, the theoretical formalism is described in Sec.~\ref{sec:theory}, where details about the $D^{(\ast)}K$ and $D^{(\ast)}_s\eta$ scattering amplitudes, the coupling of the meson-pair degrees of freedom with the bare CQM $c\bar s$ spectrum, the restoration of unitarity and the extension of the scheme to finite volume are discussed. In Sec.~\ref{sec:results} our results for the finite volume $0^+$ and $1^+$ energy levels are presented and compared with the ones reported in Ref.~\cite{Bali:2017pdv}. The properties of the $D^*_{s0}(2317)$  and $D_{s1}(2460)$ mesons, in particular their molecular content, are discussed in this section. We also compute the energy levels obtained using the unitarized NLO HMChPT amplitudes derived in Refs.~\cite{Guo:2008gp, Liu:2012zya}, and extensively compare the predictions of this latter scheme with those deduced by including a bare CQM pole. The section concludes with predictions for $DK$ S-wave scattering phase-shifts, 
and a discussion about a few aspects of the renormalization dependence of the results obtained in this work. The 
conclusions and a summary of results are presented in Sec.~\ref{sec:concl}. Finally, in Appendix~\ref{app:integrate-out}, we briefly study the relation between the NLO LECs determined in  Ref.~\cite{Liu:2012zya} and the parameters of the bare CQM  pole found in this work. 

\section{\boldmath S-wave Goldstone boson scattering off pseudoscalar and vector charm mesons: the $(S,I)=(1,0)$ sector}
\label{sec:theory}

As mentioned in the Introduction, we will extend the formalism of Ref.~\cite{Albaladejo:2016ztm} for the $\bar{B}^{(\ast)}K$ and $b\bar{s}$ system to the charm sector. We will briefly review here the most relevant aspects, paying attention to the  inclusion of the $D_s^{(\ast)}\eta$ channels, whose counterparts were not considered in Ref.~\cite{Albaladejo:2016ztm}.  

\subsection{Interactions}

We are interested in the  S-wave Goldstone boson ($K^+$, $K^0$ and $\eta$) scattering off charm mesons %
$ P^{(\ast)}_a\equiv (D^{0(\ast)},D^{+(\ast)},D_s^{+(\ast)})$. We will generically refer as $\phi$ to the former mesons and $P^{(*)}$ to the latter ones.   The heavy-light charm mesons are described in terms of HQSS matrix field $H_a$,
\begin{align}\label{eq:HQSSfields1}
\begin{split}
 H_a&=\frac{1+\slashed{v}}{2}\left(P^{\ast}_{a\,\mu}\gamma^\mu-P_a\gamma_5\right),
\end{split} 
\end{align}
with $v^\mu$ the four-velocity of the heavy mesons, and $a$ a light-flavor SU(3) index. The HQSS field  combines the isospin doublet and singlet of pseudoscalar
heavy-mesons $P_a^{(c)}=\left(c\bar u,c\bar d, c\bar s\right)$ fields and their vector HQSS partners $P_a^{\ast(c)}$. Transformations of these fields under parity and  heavy
spin and chiral rotations can be found for instance in Ref.~\cite{Albaladejo:2016ztm}. The Weinberg-Tomozawa 
Lagrangian (WTL)  describes the S-wave $P^{(*)}\phi$ chiral interaction at LO,  and it reads~\cite{Grinstein:1992qt,Wise:1992hn,Burdman:1992gh,Yan:1992gz},
\begin{align} \label{eq:Otransition1}
 \mathcal{L}_\chi=\frac{i}{2} {\rm Tr}\left(\bar H^a H_b
 v^\mu\left[\xi^\dagger\partial_\mu\xi+\xi\partial_\mu\xi^\dagger\right]_a^b\right),
\end{align}
with $\bar{H}^a=\gamma^0 H^{\dagger}_a \gamma^0$ the hermitian conjugate  of $H_a$,  and $\xi$ the Goldstone-boson matrix field given by~\cite{Wise:1992hn}, 
\begin{align}
 \xi &= {\exp}\left(i\frac{\widetilde M}{\sqrt{2}f}\right),\label{eq:xi_lightpseudoscalars}
\end{align}
with normalization $f\sim 93$ MeV and the matrix $\widetilde M$ reads,
\begin{align}\label{eq:matrix-of-goldstones}
 	\widetilde M=\begin{pmatrix}
    \frac{\pi^0}{\sqrt{2}} + \frac{\eta_8}{\sqrt{6}} + \frac{\eta_1}{\sqrt{3}} & \pi^+ & K^+\\
    \pi^- & -\frac{\pi^0}{\sqrt{2}} + \frac{\eta_8}{\sqrt{6}} + \frac{\eta_1}{\sqrt{3}}& K^0 \\
    K^- & \bar{K}^0 & -\sqrt{\frac{2}{3}}\eta_8+\frac{\eta_1}{\sqrt 3}\\    
    \end{pmatrix},
 \end{align}
where we explicitly consider the $\eta_1$ and $\eta_{8}$ unflavored SU(3) states. Note that in our normalization the heavy-light meson
field, $H$, has dimensions of $E^{3/2}$ (see
Ref.~\cite{Manohar:2000dt} for details). This is because we use a
non-relativistic normalization for the heavy mesons, which differs
from the traditional relativistic one by a factor $\sqrt{M_H}$.
 On the other hand, within the HQSS formalism, the even parity
CQM bare $c\bar q $ states, associated to the $j_{\bar  q}^\pi=\tfrac{1}{2}^+$ HQSS doublet, are described by the matrix field
$J_a$~\cite{Falk:1991nq},
\begin{eqnarray}\label{eq:HQSSfields2}
 J_a&=&\frac{1+\slashed{v}}{2}\left(Y^{\ast}_{a\,\mu}\gamma_5\gamma^\mu+Y_a\right), 
\end{eqnarray}
with $v^\mu Y^{\ast}_{a\,\mu}=0$. The $Y_a$ and $Y_a^{\ast}$ fields respectively annihilate the $0^+$ and $1^+$ bare states belonging to the $\tfrac{1}{2}^+$  doublet. Since in this work we will be interested in  the $(S,I)=(1,0)$ sector, in what follows and for simplicity we will denote $Y^{(\ast)}_{c\bar{s}}$ as $Y^{(\ast)}$. The mass of CQM bare states, $\mbare$, is a renormalization-dependent parameter of the scheme~\cite{Cincioglu:2016fkm}  that will be discussed below. At LO in the heavy quark expansion, there exists only one term invariant under Lorentz, parity, chiral and heavy quark spin transformations~\cite{Albaladejo:2016ztm},
\begin{equation} \label{eq:Otransition2}
 \mathcal{L}=\frac{ic}{2} {\rm Tr}\left(\bar H^a J_b
 \gamma^\mu\gamma_5\left[\xi^\dagger\partial_\mu\xi-\xi\partial_\mu\xi^\dagger\right]_a^b\right)+h.c.~,
\end{equation}
where $c$ is a dimensionless undetermined LEC that controls the strength of the vertex. This LEC, though it depends on the orbital angular momentum and radial quantum numbers of the CQM state, is in principle independent of the spin of the quark-model heavy-light meson, and of the light SU(3) flavor structure of the vertex. Thus, up to
$\Lambda_{\rm QCD}/m_Q$ corrections, it can be used both for $J=0$ and $J=1$ in the charm and bottom sectors. Moreover, in the SU(3) limit, the same LEC governs the interplay between two-meson and quark model degrees of freedom in all isospin and strangeness channels. This LEC was found to be $c= 0.74\pm 0.05$ in Ref.~\cite{Albaladejo:2016ztm} from  a fit to the LQCD  isoscalar $b\bar s$ $0^+$ and $1^+$  energy-levels computed in Ref.~\cite{Lang:2015hza}.

Let us consider the transitions involving  pseudoscalar  ($P$) heavy-light and Goldstone ($\phi$) mesons, 
\begin{align}
 P_I  \phi_I \to P_F \phi_F,
\end{align} 
for which the tree level isoscalar\footnote{The phase convention for isospin states $|I,I_3\rangle$ used in this work is $ \bar{u}=|1/2,1/2\rangle$ and $ \bar{d}=-|1/2,1/2\rangle$, which implies  $|D^+\rangle = -\left|\frac{1}{2},+\frac{1}{2}\right\rangle$,  while the other meson 
states are defined with a positive sign in front of the $|I,I_3\rangle$ state. Moreover, we
use the order $DK$, as in Ref.~\cite{Guo:2009ct},  to construct the isoscalar amplitudes.} amplitude ($V_c$) deduced from the WTL of Eq.~\eqref{eq:Otransition1} reads~\cite{Guo:2009ct}
\begin{align}
\begin{split}
	&V_c(s,u)=-A\frac{(s-u)}{2f^2},\\
	&A=\begin{pmatrix}
    1	& \sqrt{2/3}\\
    \sqrt{2/3}	& 0\\
    \end{pmatrix},
\end{split}
\label{eq:contact-potential}
\end{align}
where  channels 1 and 2 are $D K $ and $D_s\eta$, respectively,  $s$ and $u$ are the usual Mandelstam variables and we have considered the $\eta-\eta^\prime$ ideal mixing~\cite{Bramon:1992kr}\footnote{The ideal
mixing-angle turns out to be around  $-19.5$ degrees, while the global fit to determine
the  $\eta-\eta^\prime$  mixing angle carried out in \cite{Ambrosino:2009sc} provided a
value of $-13.3(5)$ degrees. There is an abundant literature on this subject,
and more recent theoretical works (see for instance \cite{Guo:2015dha} and references therein)
found higher absolute values for the mixing-angle, and therefore closer to the ideal one. Thus for instance, the authors of Ref.~\cite{Guo:2011pa}
quoted $-16.2^{+2.8}_{-2.9}$ degrees. Fine details of the $\eta-\eta^\prime$  mixing are
irrelevant for the exploratory study carried out in this work, since, as we will see, the influence of the $D_s^{(*)}\eta$ channels on the $D_{s0}^\ast(2317)$ and $D_{s1}(2460)$ dynamics is
quite small,  and can be taken into account by  mild variations of the rest of
the LECs of the scheme. Thus, for simplicity, we find sufficiently accurate to adopt the ideal mixing scheme.
However, $\eta-\eta^\prime$  mixing fine details are certainly relevant
for studies on the number of colors and/or pion mass dependencies of the properties of these resonances, as
the one carried out in  \cite{Guo:2015dha},  where $m_\pi$ is extrapolated until values close to 700 MeV.},
\begin{align}\label{eq:eta-mixing}
	\begin{split}
		\eta_8 = \frac{2\sqrt{2}}{3}\eta-\frac{1}{3}\eta^\prime, \\
    	\eta_1 = \frac{1}{3}\eta        +\frac{2\sqrt{2}}{3}\eta^\prime.
	\end{split}    
\end{align}
After projecting into $J=0$, we replace 
\begin{eqnarray}
 \frac{(s-u)}{2} & \to &  \frac{3s^2-s\Sigma-\Delta_I\Delta_F}{4s} \label{eq:s-u}
\end{eqnarray}
with $\Sigma= (M_I^2+m_I^2+M_F^2+m_F^2)$ and $\Delta_{I(F)}= (M_{I(F)}^2-m_{I(F)}^2)$, where $M_{I(F)}$ and $m_{I(F)}$ are the masses of the initial (final) heavy-light charm  and Goldstone mesons, respectively.

The WTL leads to similar isoscalar amplitudes for the transitions involving vector ($P^*$) heavy-light mesons,
\begin{align} 
 P^*_I  \phi_I \to P^*_F \phi_F~.
\end{align} 
Indeed, one gets in this case a potential\footnote{Now channels 1 and 2 are $D^* K $ and $D^*_s\eta$, respectively.} like that of Eq.~\eqref{eq:contact-potential}, supplemented by a term $-\epsilon_I \cdot \epsilon_F$, where $\epsilon_{I(F)}$ is the polarization four-vector of the initial (final) heavy-light meson. 
This latter factor reduces, at LO,  to 1 after projecting into $J=1$ (S-wave, i.e., zero orbital angular momentum).  In summary, the amplitudes given in  Eq.~\eqref{eq:contact-potential}, together with the projection implicit 
in Eq.~\eqref{eq:s-u} provide the coupled-channel contact potential, $V_c(s)$, both in the $0^+$ and $1^+$ sectors. 

Next, we consider  transitions between  CQM bare  states [$Y^{(\ast)}$] and $P^{(\ast)}\phi$ meson pairs, 
$ Y^{(\ast)} \to P^{(\ast)}\phi$. From Eq. \eqref{eq:Otransition2} we find 
\begin{align}	
\begin{split}
&V_{c\bar{s}}(s)=-\tilde{A}\frac{ic}{f}\sqrt{M\mbare}\frac{s+m^2-M^2}{\sqrt{s}},\\
&\tilde{A}=\begin{pmatrix}
    1\\
    \frac{1}{\sqrt{6}}\\
    \end{pmatrix},
\end{split}
\label{eq:CQM-to-mesons-transitions}
\end{align}
with $M$ an $m$ the masses of the $P^{(\ast)}$ and $\phi$ mesons, respectively.

Note that, here, by  bare mass, we mean the mass of the CQM states when the LEC
  $c$ is set to zero, and thus it is not a physical observable. In the
  sector studied in this work, the coupling 
 to the $ P^{(*)} \phi$ meson pairs renormalizes this
  bare mass, as we will discuss below. Since, in the effective theory,
  the ultraviolet (UV) regulator is finite, the difference between the bare and the
  physical resonance masses is a finite renormalization. This shift
  depends on the UV regulator since the bare mass itself depends on
  the renormalization scheme.  The value of the bare mass, which is
  thus a free parameter, can either be indirectly fitted to
  experimental observations, or obtained from schemes that ignore the
  coupling to the mesons, such as some
  CQMs. In this latter case, the issue certainly
  would be to set the UV regulator to match the quark model and the
  HMChPT approaches~\cite{Cincioglu:2016fkm}.
 
 The vertices in Eq.~\eqref{eq:CQM-to-mesons-transitions} can be used to compute the contribution [$V_\text{ex}(s)$] to $P^{(\ast)}\phi$ scattering
via the exchange of intermediate even-parity charmed-strange mesons. It is given by~\cite{Cincioglu:2016fkm, Albaladejo:2016ztm}
\begin{equation}
 \label{eq:exchange}
 V_\text{ex}(s)=\frac{V_{c\bar s}(s)V_{c\bar s}^\dagger(s)}{s-(\mbare)^2}~.
\end{equation}
Finally, the full effective potential, $V(s)$, consistent with HQSS is given by 
\begin{equation}\label{eq:fullpot}
V(s) = V_\text{c}(s) + V_\text{ex}(s),
\end{equation}
that incorporates the interplay between meson-pairs and CQM degrees of freedom in the $P^{(\ast)}\phi$ dynamics~\cite{Cincioglu:2016fkm}.  

Note that  $V_\text{c}(s)$ is obtained from $V_\text{c}(s,u)$ in Eq.~\eqref{eq:contact-potential} using the projection implicit in Eq.~\eqref{eq:s-u}. The LO potentials, $ V_\text{c}(s)$ and $ V_\text{ex}(s)$, are identical  in the heavy-quark limit in both the $0^+$ and $1^+$ sectors. Here, we will include some HQSS breaking corrections stemming  from the mass difference between pseudoscalar and vector  heavy-light mesons and the masses of scalar and axial CQM bare states.  The scheme derived here is similar to that followed in Ref.~\cite{Albaladejo:2016ztm}, though here the $\Dw \eta$ channel has been also included, while the counterpart of this channel in the bottom-strange sector was not considered in the latter work.
\subsection{Unitarity in coupled-channels}

Elastic unitarity in coupled channels is restored by solving for each $J^\pi$ sector a BSE, using as kernel the HQSS effective potential of Eq.~\eqref{eq:fullpot}. The BSE is solved  within the so-called on-shell approximation~\cite{Nieves:1999bx} and using a Gaussian cut-off, $\Lambda$, to regularize/renormalize its UV behaviour.  Thus, the unitary scattering amplitude $T(s)$ in coupled channels is obtained from the matrix equation
\begin{equation}\label{eq:InvTMatrix}
 T^{-1}(s)=V^{-1}(s)-G(s),
\end{equation}
where $G(s)$ is the regularized ($\Lambda$) two-meson loop function. Normalizations are fixed thanks to the relation between the $S$ and $T-$matrices, and the relation of the latter matrix with phase-shifts [$\delta(s)$] and inelasticities [$\eta(s)$]. Namely, we use
\begin{eqnarray}
S(s) &=& \mathbb{I}- \frac{i}{8\pi} \Sigma^\frac12 (s) T(s) \Sigma^\frac12 (s) \nonumber \\
 \eta(s) e^{2i\delta_a(s)} &=& 1-i\frac{\Sigma_{a}}{8\pi}T_{aa}(s),\quad a=1,2~,
	\label{eq:T-matrix-parametrization}
\end{eqnarray}
with $\Sigma(s) = \text{diag} \left( \sigma_1(s), \sigma_2(s) \right)$ and  the function $\sigma_{i}(s)$ is defined as:
\begin{equation} 
\sigma_{i}(s)=\frac{\lambda^{1/2}(s,M_a^2,m_b^2)}{s}\Theta\left[s-(M_a+m_b)^2\right]\quad i=1,2~,
\end{equation}
where $\lambda(x,y,z)=x^2+y^2+z^2-2xy-2xz-2yz$ and $(a,b)= (D^{(*)},K)$ and ($D^{(*)}_s,\eta)$ for $i=1$ and $2$, respectively. Thus, unitarity in coupled-channels can be expressed as
\begin{equation}
 {\rm Im} T^{-1}(s+i \epsilon) = - {\rm Im} G(s+i \epsilon)= \frac{\Sigma(s+i \epsilon)}{16\pi}~.
\end{equation}
The expression for $G$ was given in Eqs.~(14) and (15) of Ref.~\cite{Albaladejo:2016ztm} for the  
$B^{(\ast)}K$ pair, and here in coupled channels it is trivially modified to compute any of the elements of the diagonal matrix $G(s)={\rm diag}\left[G_{\Dx K}(s),G_{\Dw \eta}(s)\right]$.  
\subsection{ CQM states}\label{sec:inclusionCQM}
We have seen that through the coupling of the CQM $c\bar{s}$ and the $P^{(*)}\phi$ degrees of freedom, the effective
interaction incorporates a term ($V_\text{ex}(s)$) driven by the exchange of bare CQM states ($Y^{(\ast)}$). Such a term introduces a pole in the two-meson tree-level amplitudes, Eq.~\eqref{eq:fullpot}, located at the bare mass value, $\sqrt{s}=\mbare$. As already mentioned, it should be interpreted as the mass of the CQM state in the limit of vanishing coupling to the $P^{(*)}\phi$ meson-pairs ($c\to 0$), and therefore it is not an observable.  
The interaction with the meson cloud \textit{dresses} the CQM state through loops [Eq.~\eqref{eq:InvTMatrix}], renormalizing its mass, and the dressed  state might also acquire   a finite width, when it is located above threshold.  
At energies far enough $\mbare$, the contribution of $V_{\rm ex}$ can be regarded as small contact interaction  
that can be accounted for by means of a LEC. However, the exchange contribution becomes more important for higher energies approaching $\mbare$, and its energy dependence might then not be safely ignored. 

A priori, the value of $\mbare$ is a free parameter of the present approach, and moreover it should depend on the renormalization scheme~\cite{Cincioglu:2016fkm}. 
We will take  predictions from quenched CQMs, which in principle do not include couplings with nearby meson-meson channels. In the  $J^\pi=0^+$ sector, quark models predict, in general, $c\bar{s}$  bare masses well above $M_D+m_K$, which lead to  attractive $V_\text{ex}$ exchange interactions at $DK$ threshold, which might help in forming the $\Dvd$ resonance. 
Most quark models predict masses in the range of $2.45-2.51$ GeV for the $1^3P_0$ $c \bar s$ state~\cite{Godfrey:1986wj,Ebert:1997nk,
Lahde:1999ih,DiPierro:2001dwf}, significantly far from the $\Dvd$ experimental mass. With the aim of improving these predictions, some other models~\cite{Gupta:1994mw,Lakhina:2006fy}  
incorporated a one-loop correction diagram to the one-gluon exchange (OGE) potential, adding a spin-dependent  term to the quark-antiquark potential which affects mesons with different flavor quarks, such as the $c\bar s$ mesons. In these works, it is shown that this correction is rather small except 
for the $0^+$ sector, where large shifts are found and the mass of the CQM $1^3P_0$ state is significantly lowered ($\sim 100 $ MeV). Indeed, it is found in the $2.35$-$2.38\ \text{GeV}$ region---closer to but still above the experimental $\Dvd$ mass. Nevertheless, these predicted states would be still above the $DK$ threshold and their width would be large due to the decay into final $DK$, and difficult to reconcile with the experiment that currently provides an upper bound of a few MeV for its total width,\footnote{Being the $\Dvd$ resonance located below the $DK$ threshold, all its hadronic decays are suppressed by isospin symmetry. Depending on different dynamical assumptions, decay widths, within molecular schemes,  from 10 keV to more than 100 keV have been predicted~\cite{Lutz:2007sk, Cleven:2014oka, Guo:2018kno}.} as mentioned in the Introduction. Moreover, quark models including these modified OGE potential will still face difficulties to describe the $J^\pi=1^+$ sector, where these corrections are 
quite small, 
and the 
experimental mass pattern is similar to that found in the isoscalar-scalar sector. 

One should bear also in mind, as the one-loop OGE correction brings the bare state closer to the $DK$ threshold, the interplay between the two-meson channel and the CQM degrees of freedom might have a major impact on the description of the resonance properties and LQCD energy-levels.

Thus, in this study we will explore both types of CQMs.  On one side, we will take the $\mbare$ values for the $j_{\bar q}^\pi=\tfrac{1}{2}^+$ charmed-strange meson doublet  predicted in the CQM calculations of Refs.~\cite{Segovia:2013wma,Segovia:2012yh} (Set A in Table~\ref{Table-sets}).  Such CQM is based on the assumption that the light constituent quark mass and the exchange of pseudo-Goldstone bosons arise as a consequence of the spontaneous breaking of the chiral symmetry in QCD.  Besides, the dynamics is completed with a perturbative OGE potential and a non-perturbative screened confining interaction~\cite{Vijande:2004he, Segovia:2013wma}.   On the other hand, another set of bare masses (Set B in Table~\ref{Table-sets}) will be employed, predicted within the same CQM but
supplemented by the  one-loop OGE corrections derived in Ref.~\cite{Gupta:1994mw} (see Ref.~\cite{Segovia:2012yh} for further details).

In contrast to the bottom sector studied in Ref.~\cite{Albaladejo:2016ztm}, we see in Table~\ref{Table-sets} how the values of the bare masses lie close to the $\Dw \eta$ threshold and, consequently, this channel has been incorporated to the formalism.
\begin{table}[htb!]
	\begin{tabular}{ccccc}
    \hline\hline
           & [Set A] & [Set B] \\
   $J^\pi$ &$\mbare$ &$\mbare$  &  $(M_{\Dx}+m_K) $ & $(M_{\Dw}+m_\eta)$\\
   \hline
   $0^+$ &$ 2510.7$ & $ 2382.9$ & $2362.8$ &$2516.1$\\
   $1^+$ &$2593.1$  & $2569.7$ & $2504.2$ &$2660.0$\\
   $1^+ - 0^+$ & $82.4$ & $186.8$ & & \\ \hline\hline
    \end{tabular}
	\caption{The two sets of $\mbare$ CQM bare masses  and the $\Dx K$ and $\Dw \eta$ averaged threshold energies (in MeV units) used in this work, taken from Ref.~\cite{Segovia:2012yh}. In addition, we take an average kaon mass of $m_K=495.6$ MeV.}
    \label{Table-sets}    
\end{table} 

\subsection{\label{sec:NLO} NLO corrections}

We will also use the  $\mathcal{O}(p^2)$ (NLO) HMChPT amplitudes derived in Ref.~\cite{Guo:2008gp}, with LECs  determined from  the
lattice calculation in Ref.~\cite{Liu:2012zya} of the S-wave scattering lengths in several $(S,I)$ sectors. After unitarization, the scheme provides an accurate description of the $P\phi$ interactions in coupled channels. For instance, as it is shown in Ref.~\cite{Albaladejo:2016lbb} and already mentioned in the Introduction, the finite volume energy-levels in the $(S,I)=(0,1/2)$ channel calculated with the unitarized $\mathcal{O}(p^2)$ amplitudes, without adjusting any parameter, are in an excellent agreement with those recently reported by  the Hadron Spectrum Collaboration~\cite{Moir:2016srx}.  These chiral amplitudes predict the existence of two scalar broad resonances,  instead of only one,  with masses around 2.1 and 2.45 GeV, respectively~\cite{Albaladejo:2016lbb}. The lower pole would form, together with  the $D^\ast_{s0}(2317)$ (also correctly predicted 
in \cite{Albaladejo:2016lbb}), a SU(3) $\mathbf{\overline{3}}$ multiplet. This scheme provides also an excellent description of the  recent 
LHCb data~\cite{Aaij:2016fma} on the $B^- \to D^+\pi^-\pi^-$ final states and give solution of a number of puzzles~\cite{Du:2017zvv}. Moreover, these NLO unitarized amplitudes have been used in Ref.~\cite{Yao:2018tqn}, as input of coupled-channel Muskhelishvili-Omn\`es integral equations, whose solutions  produced  scalar form factors of the semileptonic heavy meson decays $D\to\pi \bar \ell \nu_\ell$, $D\to \bar{K} \bar \ell \nu_\ell$, $\bar{B}\to \pi  \ell \bar\nu_\ell$ and $\bar{B}_s\to K  \ell \bar\nu_\ell$ in good agreement with LQCD and light-cone sum rule predictions.

The consideration of the contributions from explicit exchanges of CQM states,  in addition to the LO  HMChPT amplitudes, contemplated in the previous subsections provides a different perspective to 
the physiognomy of the $P^{(\ast)}\phi$ interactions. Thus, it is worth discussing if there exists an energy-regime where the LO \& CQM scheme might mimic the NLO amplitudes, and when the CQM effects cannot be properly accounted for by the LECs that appear beyond  LO in the chiral expansion. Such study could provide further insights on the  energy range of applicability of the unitarized effective theory.
Hence, in addition to results obtained with a  LO \& CQM interaction, we will also  show  results from the  $(S,I)=(1,0)$ NLO input derived in Refs.~\cite{Guo:2008gp,Liu:2012zya},  and will compare both sets of predictions with the LQCD energy-levels reported in Ref.~\cite{Bali:2017pdv}.

\subsection{Poles, couplings and the compositeness condition for bound states }
  The interplay between meson-meson and CQM $c\bar s$ states might dynamically generate new states that arise as poles of the scattering amplitudes on the complex $s-$plane. There exist two thresholds $s^{(1,2)}_+$,
\begin{align}
  s^{(1)}_\pm\equiv \left(M_{D^{(\ast)}}\pm m_K\right)^2,\\
  s^{(2)}_\pm\equiv \left(M_{\Dw}\pm m_\eta\right)^2,
  \label{eq:thresholds}
\end{align}
where, for future purposes we have also defined $s^{(1,2)}_-$. Bound states are identified as poles of the scattering amplitudes located on the real axis, below the lowest threshold, $s^{(1)}_{+}$, on the first Riemann sheet (FRS).

  Additionally, resonances are identified with poles on the second Riemann sheet (SRS) of the amplitudes, 
below the real axis and above $s^{(1)}_+$.    In our two-channel problem, Riemann sheets (RSs) are denoted as
$(\xi_1\, \xi_2)$, $\xi_i = 0,1$, and are defined in the whole complex plane through analytical continuations of the loop functions~\cite{Nieves:2001wt}:
\begin{equation}\label{eq:RS}
G_{ii}(s) \to G_{ii}(s) + i \frac{\sigma_i(s)}{8\pi}\ 
\xi_i,\quad s\in \comp~,
\end{equation}
where the cuts for $\sigma_i(s)$  go along the
real axis for $-\infty < s < s^{(i)}_- $ and $ s^{(i)}_+ < s < \infty $. The
function is chosen to be real and positive on the upper lip of the
second cut, $ s^{(i)}_+ < s < \infty $. Therefore, $\sigma_i(s)$ satisfies $0\leqslant  \sigma_i(s+{\rm i}\epsilon)=-\sigma_i(s-{\rm i}\epsilon)$ for $s^{(i)}_+\leqslant s  \in \reales~$.

With all these definitions, $(00)$ is the physical RS, while the SRS is defined by requiring continuity across the unitarity cut between its fourth quadrant and the first one of the FRS. Therefore, the definition of the SRS of the amplitudes varies below and above the highest threshold (branch point of the $T-$matrix)~\cite{Nieves:2001wt}, and it corresponds to $(10)$  or $(11)$ when the real part of $s$ is above $s^{(1)}_+$, but below  $s^{(2)}_+$, or above both thresholds, respectively (see Ref.~\cite{Nieves:2001wt} for some more details).

The mass and the width of the bound state/resonance can be found from the position of the pole
on the complex energy plane. Close to the pole, the $T$-matrix behaves
as 
\begin{equation}
\label{eq:polesinT}
T_{ij}(s) \simeq \frac{g_i g_j}{s-s_R}.
\end{equation}
The quantity $\sqrt{s_R}=M_R - i\, \Gamma_R/2$ provides  the mass ($M_R$) and the width
($\Gamma_R$) of the state, and $g_i$ is the complex coupling of the resonance to the channel $i$ that it is obtained from the residue. The residues can be used to get information on the compositeness of the bound states. 
Motivated by the Weinberg compositeness condition~\cite{Weinberg:1962hj, Weinberg:1965zz, Baru:2003qq}, the probability of finding the ${D^{(\ast)}}K$ or $\Dw \eta$ molecular component in the bound state wave function is given by ~\cite{Gamermann:2009uq, Aceti:2014ala},
\begin{equation}\label{eq:Molprob}
 P_{j}=-g_j^2\left.\frac{\partial G_j}{\partial s}\right|_{s=M_b^2}~,
\end{equation}
where $M_b$ is the bound state mass.
The energy dependence of the potential produces probabilities in Eq.~\eqref{eq:Molprob} that deviate from one.  We will restrict the discussion to the case of bound states. The evaluation of Eq.~\eqref{eq:Molprob} for resonances gives rise to complex values of $P_j$, 
loosing then a straightforward  probabilistic interpretation. (Further details can be found, for instance, in section 4.2 of Ref.~\cite{Cincioglu:2016fkm}. See also Refs.~\cite{Guo:2015daa,Oller:2017alp}.)
\subsection{Finite volume and details of the simulation of Ref.~\cite{Bali:2017pdv} \label{subsec:finite-volume}}

As mentioned in the Introduction, the simulation of Ref.~\cite{Bali:2017pdv} reported finite volume energy-levels from a high statistics study of the $D^*_{s0}(2317)$ 
and $D_{s1}(2460)$ resonances,  taking into account effects of the nearby $DK$ and $D^\ast K$ thresholds by employing appropriate four-quark operators. Six ensembles with $N_f=2$ non-perturbatively ${\cal O}(a)$ improved clover Wilson sea quarks at lattice-spacing $a=0.071$ fm were employed in Ref.~\cite{Bali:2017pdv}, covering different spatial volumes and
pion masses: linear lattice size ($L$) of $1.7\ \text{fm}$ to $4.5\ \text{fm}$ were realized for $m_\pi= 290\ \text{MeV}$ and $3.4\ \text{fm}$ and $4.5\ \text{fm}$ for an almost physical pion mass of $150\ \text{MeV}$. Thus, the work of Ref.~\cite{Bali:2017pdv} represents a clear improvement on the pioneering ones of Refs.~\cite{Mohler:2013rwa, Lang:2014yfa}, where 
an ensemble with $m_\pi=156$ MeV, at a fairly coarse lattice
spacing of $a=0.09\ \text{fm}$ and a small spatial lattice
extent of $L=2.9\ \text{fm}$ ($Lm_\pi=2.29$) were analyzed using the effective range approximation to extract infinite volume results. Thus, the use of a  finer lattice spacing in Ref.~\cite{Bali:2017pdv} is important since discretization effects can be substantial for observables involving charm quarks, while exploring the dependence on the spatial volume is needed, since contributions which are exponentially suppressed in $Lm_\pi$ (that are ignored in the L\"uscher formalism) may not be small for $Lm_\pi=2.29$. 

Nevertheless, we should  warn the reader that by
employing $N_f=2$ dynamical fermions, effects arising from strange sea quarks are omitted in Ref.~\cite{Bali:2017pdv}, with the expectation that the valence strange quark provides the dominant contribution. This seems to be the case, as can be inferred from Fig.~10 of this latter reference, where  the splittings found in Ref.~\cite{Bali:2017pdv} of the two lowest states, with the noninteracting threshold for the scalar and axial-vector channels, for $m_\pi=$ 290 MeV and $150$ MeV  and various volumes are compared to the $m_\pi=$ 156 MeV 2+1 dynamical quarks results obtained in Ref.~\cite{Lang:2014yfa} for $Lm_\pi=2.29$.  The latter single-volume splittings lie in the volume-dependence curves derived in Ref.~\cite{Bali:2017pdv}, but with significant larger errors. Firstly because the number of gauge configurations used in Ref.~\cite{Bali:2017pdv} is around one order of magnitude larger than that computed in Ref.~\cite{Lang:2014yfa}, and secondly, perhaps, as consequence of the discretization errors that should be higher in this LQCD 
simulation (see for instance the sizable breaking of Lorentz symmetry in the heavy-light meson dispersion relation given in Eq.~(2) and Table VI of Ref.~\cite{Lang:2014yfa}).

To compare with the energy-levels reported in  Ref.~\cite{Bali:2017pdv}, we consider our scheme, based on unitarized HMChPT and the contribution of CQM states,  in a cubic box of side $L$, and periodic boundary conditions for the fields. The three momentum is quantized $\vec{q} = \frac{2\pi}{L}\vec{n}$ ($\vec{n}\in \mathbbm{Z}^3$). The integrals in the loop functions $G(s)$ are replaced by their finite volume versions, $\widetilde{G}(s,L)={\rm diag}\left[\widetilde{G}_{\Dx K}(s,L),\widetilde{G}_{\Dw \eta}(s,L)\right]$~\cite{Doring:2011vk,Albaladejo:2013aka}, involving the sum over all possible $\vec{q}$,
\begin{align}\label{eq:Gfv}
\begin{split}
\widetilde{G}_j(s,L) & = \frac{1}{L^3}\sum_{\vec{n}\in \mathbbm{Z}^3}
\frac{\Omega(\vec{q}\,) e^{-2(\vec{q}^{\,2}-\vec{k}^{\,2})/\Lambda^2}}{s-\left[\omega_{P^{(*)}_j}(\vec{q}\,)+\omega_{\phi_j}(\vec{q}\,)\right]^2},\\
\Omega(\vec{q}\,)      & =  \frac{\omega_{P^{(*)}_j}+\omega_{\phi_j}}{2\omega_{P^{(*)}_j}\omega_{\phi_j}},
\end{split}
\end{align}
where $ \omega_{i}(\vec{q}\,) = \left({\cal M}_i^2+\vec{q}^{\,2}\right)^\frac12$, with ${\cal M}_i$ the mass of the heavy-light meson or the Goldstone boson  for $i=P^{(*)}$ or $\phi$, respectively. (We have adopted relativistic dispersion relations as in Ref. \cite{Bali:2017pdv}.) 
Up to the order considered in this work, there are no finite volume corrections to the potential, so the full volume dependence is carried by the loop function $\widetilde{G}$  defined in a finite box.
The $T$-matrix in finite volume, $\widetilde{T}(s,L)$, is given by,
\begin{equation}
\widetilde{T}^{-1}(s,L) = V^{-1}(s) - \widetilde{G}(s,L)~.
\end{equation}
Thus, the energy-levels $E(L)$ ($s=E^2$,~$E\in \mathbbm{R}$) are computed from the poles of $\widetilde{T}(s,L)$ for each size of the box. The spectrum becomes discrete, with levels that, in principle, can be associated to 
two-meson $(P^{(*)}\phi)$ scattering states. In the non-interacting case, the free energies, $E^{\rm free}_{\Dx K}(\vec{q}\,)$ and $E^{\rm free}_{\Dw \eta}(\vec{q}\,)$ are recovered,
\begin{align}
\begin{split}
	\label{eq:free-energies}
	E^{\rm free}_{j}= \omega_{P^{(*)}_j}(\vec{q}\,) + \omega_{\phi_j}(\vec{q}\,) .
\end{split}
\end{align}
 Hence, the continuous volume dependent curves that will be presented below  
 are essentially the L\"uscher curves obtained from the phase shift by solving
\begin{equation}
\delta(q) + \phi(\hat q) = n\pi \label{eq:luscher}
\end{equation}
with $\hat q=q L/2\pi$ and $\phi(\hat q)$ determined by the L\"uscher function (see Eq.~(6.13) of Ref.~\cite{Luscher:1990ux}).

On the other hand, for a proper comparison with the results of Ref.~\cite{Bali:2017pdv}, it is necessary to use the lattice meson masses obtained in that simulation. That work reported two different sets of results that  correspond to  two different pion masses used to compute the energy-levels.  We will label here the two sets of LQCD levels by Ensembles I and II, for $m_\pi=290 $  MeV and 150 MeV, respectively. We take the $\pi$, $K$, $D$ and $D^*$ masses, for the different box-sizes considered in each of the ensembles, from Table I of Ref.~\cite{Bali:2017pdv}.  Besides, 
the $0^-$ and $1^-$  $D_s^{(\ast)}$ masses for the Ensemble II are taken from Table VII ($L/a=64$) of Ref.~\cite{Bali:2017pdv}, while for Ensemble I the masses of the  charmed-strange heavy-light mesons have been obtained using the values of $(m_{0^+}-m_{0^-})$ and $(m_{1^+}-m_{1^-})$ displayed in Fig.~13 of the latter reference, and taking for the $m_{0^+}$ and $m_{1^+}$ masses the  $L/a=64$ values reported in Table III of the same work (Ref.~\cite{Bali:2017pdv}). Hence, we neglect any dependence on $L$ in the masses of $0^-$ and $1^-$  $c\bar s$ ground states.\footnote{It is not straightforward to extract the masses of the $0^-$ and $1^-$ $D_s^{(*)}$ mesons, for each of the pion masses and volumes studied in Ref.~\cite{Bali:2017pdv}, from the results reported in that work. Note, however, that volume effects in the mases of these mesons are expected  to be even smaller than for the $D-$meson, and that these masses enter only through the small effects originated from the coupled-channels dynamics, when the 
$D_s^{(
*)}\eta$ components are considered.} The masses used in this work are compiled in Table~\ref{tab:LQCD}. 

The mass of the $\eta$ meson is not reported in Ref.~\cite{Bali:2017pdv} either, so we estimated its value, as function of the volume and pion mass,  using the Okubo mass formula~\cite{Okubo:1961jc} as follows
 \begin{align}
  m_\eta^{\rm lat}= m_\eta^{\rm exp} + m_8^{\rm lat} - m_8^{\rm exp}\,,
 \end{align}
with $ m_8^2= 4(m_K^2-m_\pi^2)/3$.  Finally for unphysical pion masses, the CQM bare masses are also corrected using the difference between the experimental and lattice spin-averaged $D_s$ masses, 

\begin{align}
\mbare^{\rm lat}  =  \mbare + \bar{M}^{\rm lat}-\bar{M}^{\rm exp},
\end{align}
with $\bar{M}=(M_{D_s}+3M_{\Dw})/4$.
\begin{table}
\caption{\label{tab:LQCD} LQCD masses (in MeV units) of the $0^-$ and $1^-$ $D_s^{(\ast)}$ mesons used in this work. They have been obtained from Tables III, VII and Fig.~13 of Ref.~\cite{Bali:2017pdv} [we always use the results corresponding to the highest volume ($L/a=64 $); see discussion in the text for details]. We also give experimental masses, taken from the  Review of Particle Physics~\cite{Olive:2016xmw}, and the difference between the experimental and lattice spin-averaged $D_s$ masses.}
 \begin{tabular}{c|c|c|c}
 \hline\hline
 & Ensemble I &  Ensemble II & Exp\\
 \hline
 $M_{D_s}$            & 1978(4)    & 1976.9(2) & 1968.28\\
 $M_{D_s^\ast}$       & 2100(7)    & 2094.9(7) & 2112.1\\
 $\bar M^{\rm exp} - \bar M^{\rm lat}$     & 5(5) & 9.6(7) \\
 \hline\hline
 \end{tabular}
\end{table}

Uncertainties in all input lattice mesons masses (Table I of Ref.~\cite{Bali:2017pdv} and Table~\ref{tab:LQCD} of this work), as well as in $\mbare^{\rm lat}$, are taken into account in the error budget of our final results, as we will detail in the next section.

To end this subsection, we would like to stress that considering the LQCD meson masses, for finite volumes and unphysical pions, it is important to set up correctly the thresholds and to properly compute the loop function in a finite box. However, the current approach will still suffer from some systematic errors, mostly because we have not considered the dependence of the Goldstone decay constant, that appear in the WTL interaction,  on the volume and the unphysical pion mass (such information is not given in Ref.~\cite{Bali:2017pdv}), and we still use its value in the infinite-volume chiral limit.\footnote{We have simplified the discussion and have focused on $f$. An accurate treatment might require  to go beyond LO in the chiral expansion, which in turn might create some problems of double counting with the contribution from CQM states. We will come back to this point below.}  Nevertheless, some of this dependence might be partially reabsorbed in the parameters fitted to the LQCD energy-levels, and we 
certainly benefit from the fact that the pions simulated in   Ref.~\cite{Bali:2017pdv} are  quite light and close to the physical one. 

\section{Results and discussion}
\label{sec:results}
\subsection{LO HMChPT+CQM analysis}\label{sec:0+results} 

\subsubsection{Fit details}

The S-wave $D^{(\ast)}K$ interaction in the LO+CQM interaction scheme depends on three, a priori, free parameters: the Gaussian cut-off $\Lambda$, the LEC $c$ (Eq.~\eqref{eq:CQM-to-mesons-transitions}) and the masses of the bare CQM $c\bar s$ state, $\mbare$. Nevertheless, we will consider different CQM meson masses in the $0^+$ and $1^+$ sectors (Sets A and B in Table~\ref{Table-sets}), as  discussed in Sec.~\ref{sec:inclusionCQM}. 

To determine the values of the LEC $c$, that controls the interplay between CQM and meson-pairs degrees of freedom,  and the Gaussian cut-off, we perform, for each set (A and B) of bare CQM masses and lattice ensembles (I and II) a combined fit to the $0^+$ and $1^+$ energy-levels, using an uncorrelated merit function defined as,
\begin{equation}
 \chi^2=\sum_i\left(\frac{E^{lat}_i-E^{th}_i}{\Delta E^{lat}_i}\right)^2\, ,
\end{equation}
where the sum spans over the $0^+$ and $1^+$ energy-levels compiled in Table III of Ref.~\cite{Bali:2017pdv}. For both $0^+$ and $1^+$ sectors, only two energy-levels for each volume are fitted. We have not considered in the fits the third $1^+$ level given in the last column of that table. It  is 
rather insensitive to the spatial volume, suggesting only a small coupling to the $D^\ast K$ threshold and, in Ref.~\cite{Bali:2017pdv},  it is identified   with the $D_{s1}(2536)$ resonance, that would presumably have a large overlap with the  $j_{\bar q}^\pi=\tfrac{3}{2}^+$ HQSS state.
\begin{figure*}
\includegraphics[height=0.40\textheight]{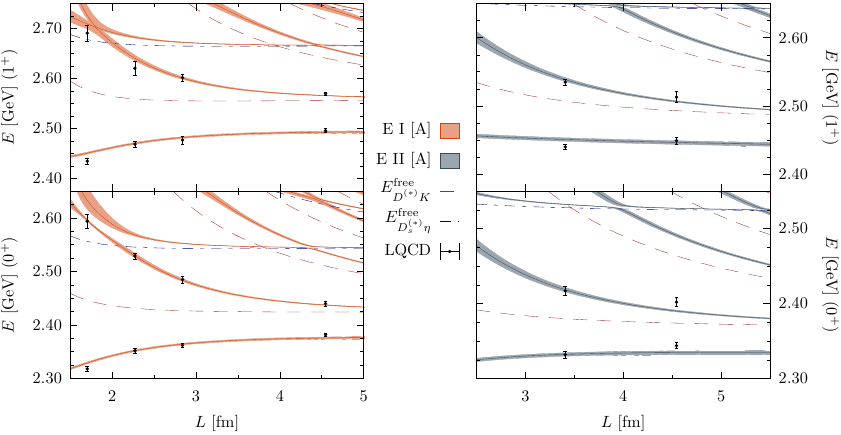}
 \caption{\label{fig:levels-1} Black points: LQCD $J^\pi=0^+$ (bottom panels) and $J^\pi=1^+$ (top panels) energy-levels, taken from Ref.~\cite{Bali:2017pdv}, 
 for different volumes and pion masses. LQCD data for Ensembles I and II are depicted in the left and right panels, respectively.
 Solid lines: Energy-levels obtained from  $0^+$ and $1^+$ combined fits to  Ensembles I and II using the Set A of CQM bare masses, as a function of the box-size, $L$ (some volume interpolated meson masses are used to compute the continuous energy-levels for values of $L$ different than  those  employed in Ref.~\cite{Bali:2017pdv}). Details of the fits and some derived quantities are collected in  Table~\ref{tab:Results}.  Statistical 68\%-confident level (CL) bands are also shown. They are calculated from the distributions obtained from  a sufficiently large number of  fits to synthetic sets of LQCD data, randomly generated assuming  that each of the LQCD energy-levels is Gaussian distributed. (Note that possible  correlations between the different energy-levels are not considered, since these are not provided in Ref.~\cite{Bali:2017pdv}.) In addition, in all  synthetic fits, the LQCD meson masses for each volume are randomly chosen as well.  For comparison, the free energies $E^{\rm free}_j$, $j=\Dx K, \Dw 
\eta$, for each volume and set of LQCD meson masses are also shown (dashed lines).}
\end{figure*}
\begin{figure*}
\includegraphics[height=0.40\textheight]{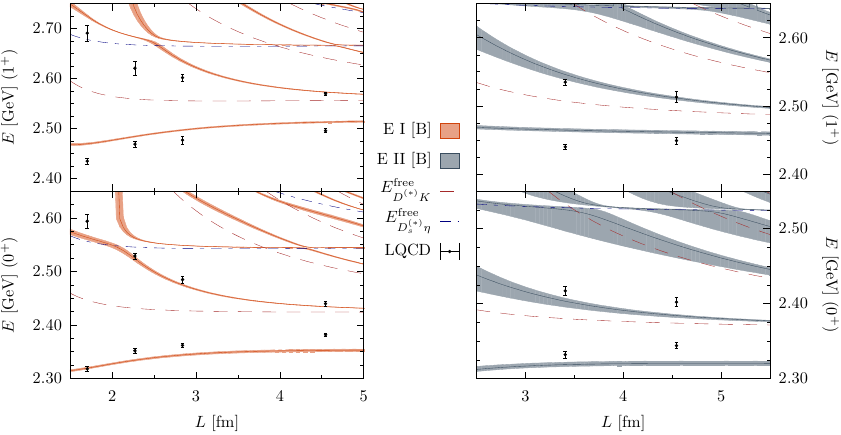}
\caption{Same as Fig.~\ref{fig:levels-1}, but in this case the predictions have been obtained using the Set B of CQM bare masses.\label{fig:levels-2}}
\end{figure*}
\begin{table*}
\caption{\label{tab:Results} Best fit LO+CQM parameters, together with infinite volume properties (masses,  $D^{(\ast)}K$ and $D^{(\ast)}_s\eta$  molecular components and couplings) of the lowest-lying $j_{\bar q}^\pi=\tfrac{1}{2}^+$ $D_s$ charm-strange meson doublet, determined from the fits to the lattice energy-levels  obtained for each of the 
two lattice pion mass ensembles (I and II) calculated in Ref.~\cite{Bali:2017pdv}, and using either Set A or B of bare CQM masses (see discussion in Sec.~\ref{sec:inclusionCQM}). S-wave isoscalar $D^{(\ast)}K$ scattering lengths ($a$) are also given, which are related to the amplitudes at threshold by $T\left[s=(M_{D^{(*)}}+m_K)^2\right]= -8\pi a \left(M_{D^{(*)}}+m_K \right)$, as in Ref.~\cite{Bali:2017pdv}. All these infinite volume quantities have been computed using physical meson masses. LQCD energy-levels and those determined in this work are shown in Figs.~\ref{fig:levels-1} and ~\ref{fig:levels-2}. 
Statistical 68\%-CL errors on the best fit parameters and derived quantities are calculated from the distributions obtained  after performing a sufficiently  large number of fits to synthetic sets of LQCD data, as explained in the caption of Fig.~\ref{fig:levels-1}. In addition,  the $c-\Lambda$ correlation coefficients are $-0.81, -0.93, 0.08$ and $-0.80$ for fits AI, AII, BI and BII, respectively. Besides, in the Set C rows,  we give the results obtained from a one-parameter (UV cutoff $\Lambda$)-fit that corresponds to an scheme where the LQCD energy-levels are described using finite-volume untarized LO HMChPT amplitudes. This is to say, the LEC $c$ is fixed to zero, and therefore the contributions to the amplitudes of the  exchange of  even-parity charmed-strange CQM mesons are neglected. The volume dependence of the $0^+$ and $1^+$ energy-levels determined within this latter scheme are shown in  Fig.~\ref{fig:levels-3} for the two lattice pion mass ensembles (I and II).}
\begin{ruledtabular}
\def\arraystretch{1.3}
\begin{tabular}{ccccccc|cccccc}
\multicolumn{7}{c|}{parameters} & \multicolumn{6}{c}{infinite volume predictions}\\
 Set & Ensemble &  $J^\pi$ & $\mbare$ & $c$ & $\Lambda$ & $\chi^2/dof$  & $M_b$  &  $P_{ D^{(\ast)}K}$ & $P_{D^{(\ast)}_s\eta}$ & $a_{ D^{(\ast)}K}$  & 
    $g_{ D^{(\ast)}K}$ & $g_{ D^{(\ast)}_s\eta}$ \\  
          &     &    &    [MeV] &     &  [MeV]& & [MeV] &  [\%] &   [\%]    & [fm] & [GeV] & [GeV] \\  
  \hline \hline
 \multirow{4}{*}{A}   & %
 \multirow{2}{*}{I}   & $0^+$ & $2511$  & %
 \multirow{2}{*}{$ 0.62\pm 0.04$} & %
 \multirow{2}{*}{$ 663^{+ 23}_{- 27 }$} & %
 \multirow{2}{*}{$ 1.8$} & %
$ 2335 \pm 2$ & $ 67\pm 1$ & $ 2.1\pm 0.2$ & $ -1.41^{+ 0.05 }_{- 0.06 }$ &  $ 10.6 \pm0.2$ & $ 5.43\pm 0.08$  \\
                       & &  $1^+$ & $2593$  &                                          &                                    &                                              & 
$ 2465\pm 2$ &  $ 57\pm 1$ & $ 1.9\pm 0.2$ & $ -1.16 ^{+ 0.03}_{- 0.04}$ & $ 12.1 ^{+ 0.3 }_{-0.2 }$ & $ 5.83 \pm 0.06$\\
\cline{2-13}
    & \multirow{2}{*}{II} & $0^+$ & $2511$  & %
 \multirow{2}{*}{$ 0.61\pm 0.09$} & %
 \multirow{2}{*}{$ 710 ^{+ 70}_{- 60 }$} & %
 \multirow{2}{*}{$3.1$} & %
$ 2331 \pm 3$ &  $ 64\pm 2 $ &  $ 2.4\pm 0.4$ &  $ -1.29^{+ 0.07}_{- 0.08} $ &  $ 10.9^{+0.4}_{-0.3}$ & $ 5.6\pm 0.1$ \\
                       & &  $1^+$ & $2593$  &                                          &                             &                                              & 
$ 2460\pm 3$  &  $ 55^{+ 2}_{- 1}$ & $ 2.2^{+ 0.4 }_{- 0.3 }$ & $ -1.07^{+ 0.05}_{- 0.06}$ & $ 12.2^{+ 0.5}_{- 0.4}$ &  $ 6.0\pm 0.1$\\
  \hline \hline
 \multirow{4}{*}{B}  &%
 \multirow{2}{*}{I}  &$0^+$ & $2383$  &%
 \multirow{2}{*}{$ 0.71\pm 0.01$ } & %
 \multirow{2}{*}{$ 426\pm 14$ } &%
 \multirow{2}{*}{$ 18.6$} & %
$ 2330 \pm2$ & $ 51\pm 1$ &  $ 0.51\pm 0.06$& $ -1.36 \pm 0.05$  & $ 11.8^{+ 0.1}_{-0.2}$ & $ 4.95\pm 0.1$\\
                        & &  $1^+$ & $2570$ &                                   &                               &                                              & 
$ 2485\pm 2$ & $ 67\pm 1$ &  $ 0.51\pm 0.07$& $ -1.79 \pm 0.09$  & $ 11.0^{+ 0.2}_{-0.3}$ &    $ 4.9\pm 0.1$ \\
\cline{2-13}
  & \multirow{2}{*}{II}  &  $0^+$ & $2383$  &  %
    \multirow{2}{*}{$ 0.57 ^{+ 0.07 }_{- 0.08 }$ } & %
    \multirow{2}{*}{$ 580 ^{+ 80}_{- 50}$} & %
    \multirow{2}{*}{$18.0$}& %
$ 2320\pm4$ & $ 45^{+2}_{-1}$& $ 1.2^{+ 0.3 }_{- 0.2 }$ & $ -1.04^{+ 0.09 }_{- 0.08 }$ & $ 11.0^{+0.5}_{-0.6}$ & $ 5.0\pm0.1$\\
                       & &  $1^+$ & $2570$  &                                   &                            &                                              & 
$ 2477\pm4$ & $ 60\pm 2$ & $ 1.3 ^{+ 0.4 }_{- 0.2 }$ &  $ -1.40\pm 0.1$ &  $ 11.1^{+0.4}_{-0.5}$ & $ 5.3^{+ 0.2}_{- 0.1}$\\
\hline \hline
 \multirow{4}{*}{C}   &%
 \multirow{2}{*}{I} &  $0^+$ & $-$  &%
 \multirow{2}{*}{$0$ fixed} &%
 \multirow{2}{*}{$ 1141\pm 14$} &%
 \multirow{2}{*}{$ 4.5$} &%
$ 2323\pm 2$ & $ 65\pm 1$ & $ 5.67\pm 0.07$ & $ -1.10\pm 0.03$ &  $ 10.9 \pm0.1$ & $ 6.59\pm 0.04$  \\
                       & &  $1^+$ & $-$  &                                          &                                    &                                              & 
$ 2463\pm 2$ &  $ 65\pm 1$ &  $ 5.64 \pm 0.07$ & $ -1.08\pm 0.03$ &  $ 11.7\pm0.1$ & $ 7.03 \pm 0.05$\\
  \cline{2-13}
  						&%
\multirow{2}{*}{II} &  $0^+$ & $-$  &%
\multirow{2}{*}{$0$ fixed} &%
\multirow{2}{*}{$ 1140\pm 20$} &%
\multirow{2}{*}{$ 2.5$} & %
$ 2323\pm3$ & $ 65\pm 1$ & $ 5.66\pm 0.1 $ & $ -1.11 \pm 0.05$ &  $ 10.9\pm 0.2 $ & $ 6.58\pm 0.07$  \\
                       & &  $1^+$ & $-$  &                                          &                                    &                                              & 
$ 2463\pm3$ & $ 65\pm 1$ & $ 5.64\pm 0.1 $ & $ -1.08 \pm 0.04$ &  $ 11.7\pm 0.2 $ & $ 7.03\pm 0.07 $
\end{tabular}\\
\end{ruledtabular}
\end{table*}
\begin{table}[htb!]
\caption{\label{tab:Results2} Infinite volume properties of the second pole, resonance located above the $D^{(\ast)}K$ threshold, that appears in the unitarized amplitudes considered in Table~\ref{tab:Results2}. Note that now the couplings are complex in general, and we only give here the moduli.} 
\begin{ruledtabular}
\begin{tabular}{ccc|cccc}
\def\arraystretch{1.3}
  Set &  Ensemble & $J^\pi$ &  $M_R$  & $\Gamma_R$   & $|g^R_{D^{(\ast)}K}|$ & $|g^R_{ D^{(\ast)}_s\eta}|$\\  
   &   &  &  [MeV] & [MeV]  & [GeV] & [GeV]\\  
  \hline
  \multirow{4}{*}{A} & \multirow{2}{*}{I} &%
     $0^+$ & $ 2689^{+ 25 }_{- 18}$ & $ 85\pm 4$ &  $ {3.5}^{+ 0.1 }_{- 0.2 }$ & $ {3.7}\pm 0.2$  \\
& &  $1^+$ & $ 2772^{+ 24 }_{- 18}$ & $ 98\pm 5$ &  $ {4.3}^{+ 0.1 }_{- 0.2 }$ & $ {3.8}\pm 0.2$\\
  \cline{2-7}
  & \multirow{2}{*}{II} & %
      $0^+$ & $ 2684^{+ 53}_{- 45}$ & $ 85^{+ 8}_{- 13}$ & $ {3.6}^{+ 0.1 }_{- 0.5 }$ & $ {3.6}\pm 0.4$  \\
&  &  $1^+$ & $ 2767^{+ 51}_{- 44}$ & $ 98^{+10}_{- 16}$&  $ {4.4}^{+ 0.1 }_{- 0.4 }$ & $ {3.8}\pm 0.4$ \\
\hline
\multirow{4}{*}{B} & \multirow{2}{*}{I} &%
      $0^+$ & $ 2602 \pm 8 $ & $ 97\pm 3$ & $ {3.4}\pm 0.2$ & $ 4.2 \pm 0.1$\\
&  &  $1^+$ & $ 2797 \pm 7 $ & $ 91\pm 2$ & $ {2.2}\pm 0.3$ & $ 4.6 \pm 0.1$\\
  \cline{2-7}
  & \multirow{2}{*}{II} &%
      $0^+$ & $ 2527^{+ 33 }_{- 32 }$ & $ 101^{+ 7 }_{- 17 }$ & $ {4.9}^{+ 0.1 }_{- 0.3 }$ & $ {3.2}\pm 0.3$\\
&  &  $1^+$ & $ 2713^{+ 36 }_{- 35 }$ & $  88^{+ 5 }_{- 12 }$ & $ {4.5}^{+ 0.2 }_{- 0.4 }$ & $ {3.5}^{+ 0.4}_{- 0.3 }$ 
\end{tabular}
\end{ruledtabular}
\end{table}

We show in Figs.~\ref{fig:levels-1} and \ref{fig:levels-2}  the  $0^+$ and $1^+$ energy-levels obtained using sets A and B of CQM bare masses, respectively. Fitted parameters and best fit $\chi^2/dof$ values are collected in  Table~\ref{tab:Results}. 

We see that the Set A of bare CQM masses provides a fairly good description of the volume dependence of the LQCD energy-levels in both the 
$J^\pi=0^+$ and $1^+$ sectors, despite  the large
deviations from the free levels. There exists  a very mild dependence of the UV cutoff and  LEC $c$ on the pion mass, which however is not statistically significant. The $D^{(*)}_s\eta$ coupled channel effects are negligible, except perhaps for the highest levels calculated with the heaviest pion mass ensemble at the smallest of the volumes, since that threshold is located sufficiently more higher than the measured levels in Ref.~\cite{Bali:2017pdv}. (Note that $D^{(*)}_s\eta$ correlators are not considered in the LQCD study of the latter reference.) We find $c=0.61 (9)$ and $\Lambda= 710(70)$ MeV from the fit to the lightest pion mass ensemble. This parameter was also determined  in Ref.~\cite{Albaladejo:2016ztm}  from a similar analysis of  the LQCD low-lying $0^+$ and $1^+$ $B_s-$energy-levels  calculated in Ref.~\cite{Lang:2015hza}. There, it was found  $c=0.75(6)$, with an UV cutoff  in the range 620-770 MeV, which points out to some small dependence of the LEC $c$ on the heavy-flavor mass. 

On the other hand, when the Set B of $0^+$ and $1^+$ CQM bare masses are used, we find unacceptable fits, with  $\chi^2/dof$ values above 18 for both pion mass ensembles. Indeed, as can be seen in Fig.~\ref{fig:levels-2}, the set B leads to a really  poor description of the LQCD data. For the latter, the HQSS breaking corrections thus look {\it i)} compatible with those encoded in the current scheme when the Set A of bare masses is used, but {\it ii)}  much smaller than those implemented by the Set B of bare masses. The one-loop corrections~\cite{Gupta:1994mw} to the OGE potential implemented in Ref.~\cite{Segovia:2012yh} produce a $1^+-0^+$ shift of the bare CQM masses of around $190\ \text{MeV}$, while it is around only $80\ \text{MeV}$ when these corrections are neglected (see Table~\ref{Table-sets}). This is because, as we already mentioned,  this correction mostly affects the $0^+$ sector~\cite{Gupta:1994mw}.  Actually, because of the denominator in Eq.~\eqref{eq:exchange} and for fixed $c$ and $\Lambda$, the decrease of the $0^+$ bare mass produces an enhancement of the attraction close to the $DK$ threshold from the exchange of the CQM state. This effect  is much less important in the $1^+$ sector, and thus the current scheme using Set B of bare masses produces a visible tension between the predicted $0^+$  and $1^+$ levels and the LQCD data (Fig.~\ref{fig:levels-2}). The scheme tends to  overestimate (underestimate) the  attraction in the former (latter) energy-levels by around $10$-$20\ \text{MeV}$, amount significantly larger than  the errors of the LQCD data. Therefore, this study strongly disfavors the bare masses found in Ref.~\cite{Segovia:2012yh}, where the one-loop corrections derived in Ref.~\cite{Gupta:1994mw} are taken into account. These one-loop corrections to the OGE potential produced a much smaller HQSS breaking of the bare masses in the bottom sector, and thus nothing statistically meaningful could be concluded about this issue in the analysis carried out in 
Ref.~\cite{Albaladejo:2016ztm} of the  $B_s-$energy-levels  reported in Ref.~\cite{Lang:2015hza}. Due to the former discussion, we will always make reference to the results obtained from the Set A of CQM bare masses in the rest of the work.

Next, and once the parameters of the model have been fixed, we  search for poles in the FRS (bound states) and SRS (resonances) of the  isoscalar S-wave $D^{(\ast)} K$ and $D^{(\ast)}_s\eta$
amplitudes for the infinite volume case and using physical meson masses. For both sets of CQM bare masses, and in both $J^\pi=0^+$ and $1^+$ sectors, we find  a bound state (FRS) and a resonance (SRS). The masses and couplings [see Eq.~\eqref{eq:polesinT}]  of the lowest-lying $j_{\bar q}^\pi=\tfrac{1}{2}^+$ charm-strange meson doublet are compiled in Table~\ref{tab:Results}, together with the $0^+$ and $1^+$ isoscalar $D^{(\ast)}K$ scattering lengths and the probabilities of the molecular $ D^{(\ast)} K$ and $D^{(\ast)}_s\eta$ components [see Eq.~\eqref{eq:Molprob}].  The properties of the additional $0^+$ and $1^+$ states, resonances located above the $D^{(*)}K$ threshold, are compiled in Table~\ref{tab:Results2} for the different schemes presented in the previous table.
  
\subsubsection{Properties of the lowest-lying states: masses, molecular probabilities and couplings}

We first pay attention  to the mesons of the lowest-lying $j_q^\pi=\tfrac{1}{2}^+$ $D_s$ doublet. We see that the predicted  mass of the $0^+$ bound state, using Set A of CQM bare masses,  is only around $15$ MeV higher than the experimental mass of the $D_{s0}^\ast(2317)$ and 75-80 MeV smaller than the CQM bare mass, while that of the $1^+$ nicely agrees, taking into account the errors, with the experimental mass of the $D_{s1}(2460)$ state. Nevertheless, this level of discrepancy can be well  attributed to the presence of discretization errors, or some uncertainties in the determination of the mass of the charm quark in the lattice simulation. Actually, the results compiled in Table I of Ref.~\cite{Bali:2017pdv}  show discrepancies of the order of 10 MeV between the experimental masses of the $D$ and $D^*$ mesons and the LQCD ones,  determined from the lightest quark mass, which provides almost physical pion masses.   We should also note some differences (1-2 $\sigma$'s) between the $0^+$ and $1^+$ masses 
found in this work and those reported in Table VII of Ref.~\cite{Bali:2017pdv}, taken from the $m_\pi=150$ MeV and $L=64a$ ensemble. These might be due to the use here of physical meson masses, and also because the LQCD ones are accessed via the L\"uscher's relation [Eq.~\eqref{eq:luscher}] using the effective range approximation.  The approach followed here, where the two meson loop function is computed in a finite volume, the unknown LECs are determined 
from fits to the LQCD data, and  finally  poles are searched for in the infinite volume unitarized chiral amplitudes, provides a theoretically sound tool to analyze the LQCD energy-levels. A good example of the latter affirmation can be found in Ref.~\cite{Albaladejo:2016lbb}, where such approach led to the existence of two $D^\ast_0(2400)-$poles, instead of only one reported in the original lattice work of the Hadron Spectrum Collaboration~\cite{Moir:2016srx}, where the LQCD energy-levels were calculated (see also the discussion in Ref.~\cite{Du:2017zvv}, where it is emphasized how the two-pole pattern of the $D^\ast_0(2400)$, together with their SU(3) structure, provides a natural solution to a number of puzzles). 

Interestingly, we appreciate in Fig.~\ref{fig:levels-1} a quite significant dependence of the lowest-lying LQCD energy-levels on the pion mass (differences between left and right plots), as it is also evident in the results of Table IV of Ref.~\cite{Bali:2017pdv}. Thus, the  LQCD $0^+$ and $1^+$ masses reported in that table vary between 30 to 50 MeV, when the pion mass is reduced from 290 MeV down to 150 MeV. These changes are likely related to the modifications of the $DK$ and $D^{*}K$ thresholds. All of this clearly indicates that the $D_{s0}^\ast(2317)$ and $D_{s1}(2460)$ states should have a sizable molecular component, and that any CQM $c\bar s$ component in their dynamics cannot be dominant,  because it could not accommodate such visible dependence of their masses on the light quark mass, as  exhibited by the LQCD data. These findings are corroborated by the molecular probabilities collected in Table~\ref{tab:Results}. Using the modified Weinberg compositeness condition of Eq.~\eqref{eq:Molprob}, we 
derive the molecular S-wave $D^{(*)}K$ probabilities for the $D_{s0}^\ast(2317)$ and $D_{s1}(2460)$, which turn out to be around 65\% and  56\% for the scalar and axial states, respectively. On the other hand, $D^{(*)}_s \eta$ components are small for both mesons, of the order of 2\%. The LQCD studies of Ref.~\cite{Bali:2017pdv,Mohler:2013rwa, Lang:2014yfa} showed a non-zero overlap of the  energy-levels related to the $D_{s0}^\ast(2317)$ and $D_{s1}(2460)$ and meson-meson lattice  interpolating fields, but no precise information was provided on the percentage of meson-meson channels in the  wave function of these states. Only in the latest work of Ref.~\cite{Bali:2017pdv}, the compositeness probability is studied, and found to be 1 within errors (around 20-30\%) for both states.

The large molecular probabilities found in this work confirm those reported in previous works~\cite{Liu:2012zya,Torres:2014vna,Albaladejo:2016hae} that, employing also unitarized meson-meson amplitudes, had already obtained  molecular components of around 70\% for the $D_{s0}^\ast(2317)$, as mentioned in the Introduction. In what respects the $D_{s1}(2460)$, the authors of Ref.~\cite{Torres:2014vna} found a molecular probability of $0.57\pm 0.22$ also in good agreement with our findings, although with a much larger  error. The interesting and novel aspect of the present calculation is that the LO HMChPT interactions have been supplemented by those driven by the exchange of  even-parity charmed-strange CQM mesons, and thus the couplings of CQM $c\bar s$ and $P^{(*)}\phi$ degrees of freedom have been explicitly taken into account.\footnote{Other studies have done something similar (\textit{e.g.} Ref.~\cite{Torres:2014vna}) by including in the interactions a Castillejo-Dalitz-Dyson pole~\cite{Castillejo:1955ed} 
to visualize a genuine (CQM) state that couples weakly to a meson-meson component. However, those studies do not make use of the HQSS to relate the interplay between both types of degrees of freedom in the $0^+$ and $1^+$ sectors, which will be fundamental to disfavor the Set B of CQM bare masses used in the study of Ref.~\cite{Ortega:2016mms}, that will be discussed in the next paragraph, and that claimed a much smaller ($\sim\!\!30\%$) molecular components for the $D_{s0}^\ast(2317)$. Moreover, in some of these studies chiral symmetry is not fully used to constraint the $P^{(*)}\phi$ interactions, and different solutions were obtained with many sets of parameters, obviously  correlated, though the claim in Ref.~\cite{Torres:2014vna} was that the particular values of the parameters did not have a special significance, and all of them led to similar hadronic-molecular probabilities~\cite{Torres:2014vna}.}

In Ref.~\cite{Ortega:2016mms} two-meson loops and CQM bare poles are also coupled. For the latter, the values of the bare masses are the same as those used here. The $D^{(\ast)} K$ interactions are derived from the same CQM used to compute the bare states, instead of using HM$\chi$PT. The $^3 P_0$ model is employed to couple both types of degrees of freedom, and the quark model wave functions provide form-factors that regularize the meson loops. Thus, all the inputs in this approach are constrained/determined from previous studies. The masses of the $0^+$ and $1^+$ states found in Ref.~\cite{Ortega:2016mms} are about 10 and 25 MeV higher than the experimental ones, respectively.  The coupling of the CQM mesons to the $DK$ and $D^*K$ thresholds is crucial  to simultaneously lower the masses of the corresponding $D_{s0}^\ast(2317)$ and $D_{s1}(2460)$ states predicted by the naive quark model. Such effects are of the order of $60$ and $85\ \text{MeV}$ in the $0^+$ and $1^+$ sectors, respectively. However, in 
the study carried out in Ref.~\cite{Ortega:2016mms}, the one-loop corrections derived in Ref.~\cite{Gupta:1994mw} are taken into account, and they lower the predicted mass of the $D_{s0}^\ast(2317)$ by more than $100\ \text{MeV}$. At this respect, we should repeat once more that the simultaneous analysis of the $0⁺$ and $1^+$ LQCD energy-levels of Ref.~\cite{Bali:2017pdv} carried out in this work strongly disfavors such corrections. Molecular probabilities are reported in Ref.~\cite{Ortega:2016mms} to be around $33\%$ and $54\%$ for the $D_{s0}^\ast(2317)$ and $D_{s1}(2460)$, respectively. Though the latter one turns out to be in a nice agreement with our results, the former one is around twice  smaller than that found here and in previous works~\cite{Liu:2012zya, Torres:2014vna, Albaladejo:2016hae}, and it would contradict a dominant molecular picture for the $D_{s0}(2317)$. Moreover, this disparity between the molecular components in the scalar and axial states  might also question  that  the $D_{s0}^\ast(
2317)$ and $D_{s1}(2460)$ could form a HQSS doublet. Within our scheme, it is however natural to assign these two states to the $j_{\bar q}^\pi=1/2^+$ HQSS doublet, assignation that gets further support from the observation that the experimental mass splitting between these two resonances is remarkably similar to that between the $D$ and $D^*$ mesons. Furthermore, interpreting  
the $D_{s0}^\ast(2317)$ and $D_{s1}(2460)$ as $DK$ and $D^*K$ bound states, the binding energies of both states will be very similar (approximately 46 MeV versus 54 MeV).

The couplings of the $D_{s0}^\ast(2317)$ and $D_{s1}(2460)$ to $D^{(*)}K$ and $D_s^{(*)}\eta$ are also compiled in Table~\ref{tab:Results}. We see that  the coupling of both states to the latter channel, though around a factor two smaller than to $D^{(*)}K$, is not negligible.\footnote{The much larger differences found for the molecular probabilities are due, not only because it appears the square of the couplings, but also because the large distance of the $D_s^{(*)}\eta$ thresholds to the masses of the resonances.} The analysis adopted in the original LQCD work of Ref.~\cite{Bali:2017pdv} led to  $g_{ D^{(\ast)}K}$ couplings of $11.0 \pm 1.3$ GeV and  $13.8 \pm 1.3$ GeV for the $D_{s0}^\ast(2317)$ and $D_{s1}(2460)$, respectively.  These values are in good agreement with the values found in this work. We would like to stress that the clear similarities between the couplings of both resonances reinforces our conclusion that they form a HQSS doublet. Moreover, 
the $D_{s0}^\ast(2317)$ and $D_{s1}(2460)$ should be heavy-flavor partners of the $\bar B_s$ scalar and axial mesons found in Ref.~\cite{Albaladejo:2016ztm} at $5709\pm 8$ and $5755 \pm  8$ MeV, respectively. Note that the mass shift, due to the breaking of HQSS, is much smaller in the bottom sector, and it turns out to be quite similar to $(M_{\bar B^* }- M_{\bar B })$, as expected. Note that 
the predictions of Ref.~\cite{Albaladejo:2016ztm} for the bottom sector agree quite well with those found in Ref.~\cite{Altenbuchinger:2013vwa}, obtained within a covariant formulation of 
unitary chiral perturbation theory involving charm mesons.

\subsubsection{$D^{(\ast)}K$ scattering lengths}

The $D^{(\ast)}K$ scattering lengths are negative (see Table~\ref{tab:Results}), compatible with the interpretation of the  $D_{s0}^\ast(2317)$ and $D_{s1}(2460)$  as bound states. Indeed, the zero-range approximation,  $a_0 = -1/(2\mu B)^\frac12 $ [$B>0$ and $\mu$ are the $D^{(*)}K$ binding energy and reduced mass, respectively], provides already the first significant digit ($-1$ fm). This simple formula also anticipates that $|a_{DK}| > |a_{D^*K}|$. Our predictions for the scattering lengths are consistent, within errors, with previous lattice determinations~\cite{Mohler:2013rwa,Lang:2014yfa, Bali:2017pdv}, and the bulk of the small differences existing among central values can be explained in terms of the differences between binding energies. The uncertainties on our estimates are significantly smaller than those affecting 
the LQCD ones. This is in line with the previous discussion about the errors on the masses of the 
$D_{s0}^\ast(2317)$ and $D_{s1}(2460)$ resonances, and it shows, once more, that the present analysis of the LQCD energy-levels leads to more precise results than those based on the L\"uscher's relation using the effective range approximation.  

\subsubsection{Second pole: resonances}

Within the current scheme, the amplitudes include an explicit pole. It is therefore reasonable to assume that the CQM bare state does not disappear, but it gets dressed by the meson-meson interaction  moving into the complex plane. In addition to a mass shift, the new state  acquires a width since it can decay into S-wave $D^{(\ast)}K$ and $D^{(\ast)}_s\eta$ meson-pairs. The position and couplings of these  extra poles, located in the SRS of the amplitudes, are collected in Table~\ref{tab:Results2}. We see that both in the $0^+$ and $1^+$ sectors, the resonances are relatively broad ($85(4)$ and $98(5)\ \text{MeV}$), respectively) and have similar couplings\footnote{The couplings are now complex in general, and we refer to the moduli.} to $D^{(\ast)}K$ and $D^{(\ast)}_s\eta$. On the other hand, the couplings of these  resonances to $D^{(*)}K$  are around a factor of three smaller than those of the $D_{s0}^\ast(2317)$ and $D_{s1}(2460)$ states. 

The masses of these higher states are 175 (330)  and 110 (270) MeV above the $D_s^{(*)}\eta$ ($D^{(*)}K$) threshold. Thus, we should take these results with some caution, since most likely they should be affected by sizable higher order chiral corrections and higher threshold-channels corrections. In other words, they are not as theoretically robust as those concerning the
lowest-lying $D_{s0}^\ast(2317)$ and $D_{s1}(2460)$ states. As mentioned, these additional  
resonances are likely originated from the bare $c\bar s$-quark-model poles that are dressed by the $ D^{(\ast)}K$ and $ D^{(\ast)}_s\eta$ meson loops.  In that case, the bare poles have been highly renormalized, moving to significant higher masses and acquiring a significant width. We should also bear in mind that radial excitations ($2^3P_0$) of the CQM states~\cite{DiPierro:2001dwf} or $D^{(\ast)}K^{\ast}$ two-meson loops, neither of them taken into account in this study, might lie in this region of energies, then having a certain impact in the dynamics of these resonances.

\subsection{LO and NLO unitarized HMChPT analysis}\label{subsec:LOandNLO}
\subsubsection{LO HMChPT energy-levels}
\begin{figure*}
\includegraphics[height=0.40\textheight]{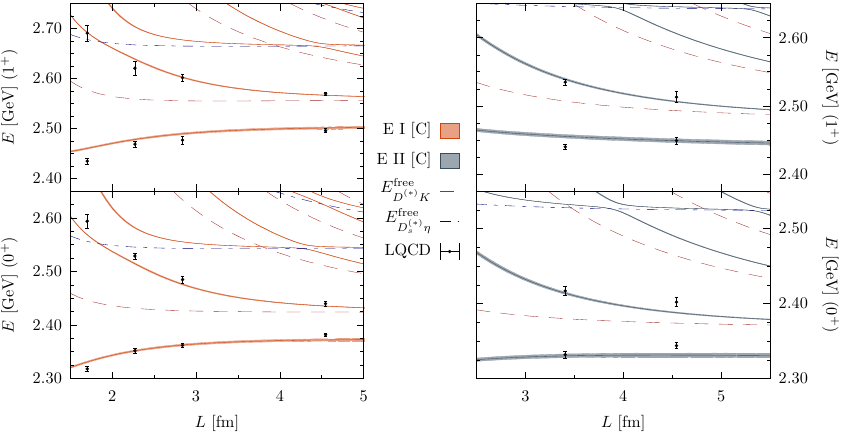}
\caption{Same as Fig.~\ref{fig:levels-1}, but in this case the predictions have been obtained after neglecting the contribution 
to the amplitudes of the  exchange of intermediate even-parity charmed-strange CQM mesons, {\it i.e.}, setting $c=0$ and evaluating the energy-levels using finite-volume unitarized LO HMChPT amplitudes. The besfit UV cutoff in this scenario and some derived quantities are given in the Set C rows of  Table~\ref{tab:Results}. 
\label{fig:levels-3}}
\end{figure*}
In addition to the results shown in the previous sections, where a CQM pole was added to the LO $D^{(\ast)}K$ interaction, 
it is enlightening to discuss whether we are able to describe the lattice data without any CQM $c\bar{s}$ contribution. To explore this scenario, we have performed an additional one-parameter (UV cutoff $\Lambda$)-fit, where the LQCD energy-levels are described using finite-volume untarized LO HMChPT amplitudes. Thus the LEC $c$ is set to zero, and consequently the contribution to the amplitudes of the  exchange of the CQM mesons vanish as well. This fit is labeled as Set C in Table~\ref{tab:Results}, and the obtained $0^+$ and $1^+$ energy-levels, as a function of the box-size, are shown in Fig.~\ref{fig:levels-3}. 

We find a quite reasonable description of the LQCD data, and the infinite volume properties of the lowest-lying states agree well with those deduced using the Set A of CQM bare masses, though molecular probabilities and couplings of the $D_{s1}(2460)$ and $D_{s0}^\ast(2317)$ are now much more similar.\footnote{We find $P_{ D^{(\ast)}K}\sim 65\%$  and  $P_{D^{(\ast)}_s\eta}\sim 6\%$ for both states, and adding the probabilities, we obtain  molecular components above 70\% in the wave-functions of both mesons. On the other hand, the higher  $D^{(\ast)}_s\eta$ channel becomes also more important in their dynamics.} Note  that the CQM exchange potential induces some HQSS breaking corrections, driven by the $0^+$ and $1^+$ $c\bar s$ bare mass-shift, and the fact that these contributions have been eliminated might explain the found pattern of probabilities and couplings. The more distinctive difference, however,  is that the UV cutoff is around 1100 MeV.  This is to say, the new UV cutoff is around 400 MeV higher 
than the values needed when the contribution of the CQM meson exchanges are kept. That reveals that higher order chiral corrections,
 previously effectively accounted for by means of the CQM pole, are not negligible. Conversely, taking into account explicitly the exchange of (bare) CQM mesons is not crucial to describe the $D_{s0}^\ast(2317)$ and $D_{s1}(2460)$ states, because  such contributions can be accommodated by appropriately modifying the finite contributions derived from short-distance physics. This is expected since the CQM bare states lie far from the latter physical states, for which the unitarity meson loops play a fundamental role.

The UV cutoff $\Lambda$ is expected to be larger than the wave number of the states ($\sim 200$ MeV) and, at the same time, small enough to prevent it from inducing large HQSS breaking corrections. From this perspective, one might think that values in the range of 1.1 GeV, comparable to the mass of the charm quark, do not seem appropriate within the spirit of an EFT based on HQSS. However, one should bear in mind that here we are using a Gaussian UV regulator, which will approximately correspond  to  a sharp-cutoff, $q_{\rm max}$, smaller by a factor\footnote{Note that the Gaussian regulator introduced in \cite{Gamermann:2009uq} and that used here are related by an extra factor $1/\sqrt{2}$.} $\sqrt{\pi/8}$~\cite{Gamermann:2009uq}. Thus, in terms of a sharp-cutoff,  we are dealing with values of the order of 700 MeV that are more acceptable from the  point of view of a HQSS EFT. 

The Set A of Gaussian regulators found in Subsec.~\ref{sec:0+results} would correspond to sharp cutoffs of the order of 400-450 MeV, which are still larger than the wave number of the $D_{s0}^\ast(2317)$ and $D_{s1}(2460)$ states. Nevertheless, we should remind here that the CQM bare masses are not observables, and they depend on the UV renormalization. Here we have fixed the CQM masses,  and thus the fitted cutoffs should be understood as those needed to effectively account for higher order chiral corrections, when these bare poles are incorporated~\cite{Cincioglu:2016fkm}. This also hints to a certain scale for which the CQM of Ref.~\cite{Ortega:2016mms} might match the chiral EFT.

\subsubsection{NLO HMChPT energy-levels}
\begin{figure*}
\includegraphics[height=0.40\textheight]{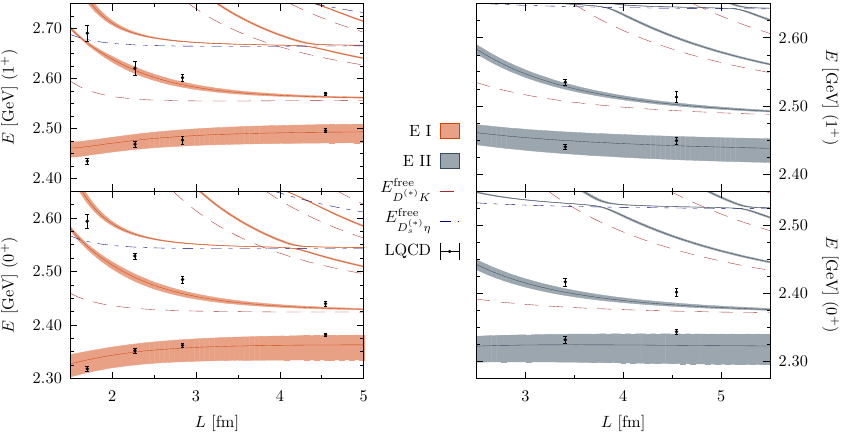}
\caption{\label{fig:levels-NLO} Volume dependence of the energy levels predicted using the unitarized NLO HMChPT amplitudes of Refs.~\cite{Liu:2012zya,Guo:2008gp} compared to the lattice results of Ref.~\cite{Bali:2017pdv}. Exchanges of intermediate CQM mesons are not included, and the distribution of panels is the same as in  Fig~\ref{fig:levels-1}. There are no fitted parameters involved in these predictions since all LECs that appear in the definition of the chiral amplitudes  were determined from  the
lattice calculation \cite{Liu:2012zya} of the S-wave scattering lengths in several $(S,I)$ sectors. The 68\% CL uncertainty bands depicted in the plots are inherited from  the errors on  the LECs quoted in Ref.~\cite{Liu:2012zya}.}
\end{figure*}
Within this context, it seems appropriate to calculate the energy-levels obtained from the unitarized HMChPT NLO amplitudes~\cite{Guo:2008gp} described in Subsec.~\ref{sec:NLO}. As in the previous subsection, the exchanges of CQM bare poles are not included. Indeed, as we have discussed above,  considering such contributions, together with the NLO corrections,  might lead to some double-counting problems. In Appendix~\ref{app:integrate-out}, we briefly study the relation between the NLO LECs determined in  Ref.~\cite{Liu:2012zya} and the parameters of the bare CQM  pole found in this work. 

The UV divergences are renormalized in Ref.~\cite{Guo:2008gp} by using subtracted meson loop functions instead of a sharp-cutoff. However, both schemes can be related (see for instance the discussion in Appendix A of Ref.~\cite{Yao:2018tqn}), and the subtraction constants determined in \cite{Liu:2012zya} correspond, in good approximation, to a sharp-cutoff  $q_{\rm max}=0.72^{+0.07}_{-0.06}$ GeV, similar to the values mentioned in the previous discussion. 

Results are shown in Fig.~\ref{fig:levels-NLO}, where we see that this scheme provides a more than acceptable description  of the $0^+$ and $1^+$ LQCD data, 
for both pion mass ensembles, despite having set all LECs to the values determined in Ref.~\cite{Liu:2012zya}. We emphasize that, therefore, the energy levels shown in the different panels of the figure are predictions and do not imply any fine adjustment of any type of parameter.\footnote{We should mention that the untarized NLO HMChPT description of the LQCD energy levels shown in Fig.~\ref{fig:levels-NLO}  might be improved by allowing for a slight variation of the LECs, which could, for example, effectively  account for some discretization/finite volume errors etc. Note that, in addition,  these systematic errors could be also different to those affecting the lattice study of scattering lengths carried out in Ref.~\cite{Liu:2012zya}. Nevertheless, the second levels are quite far from the thresholds, and one might need to explicitly include higher order chiral corrections. Otherwise, the re-fitted NLO LECs would be biased, since they   would effectively account for those contributions.   All of this is 
beyond the scope of the present work. } We find this remarkable, and together with the  similar good description found in Ref.~\cite{Albaladejo:2016lbb} of the $(S,I)=(0,1/2)$ LQCD low-lying energy levels calculated in Ref.~\cite{Moir:2016srx}, provides a great support for the finite-volume amplitudes obtained after unitarizing the NLO HMChPT amplitudes derived in Refs.~\cite{Guo:2008gp,Liu:2012zya}. Hence, the $D^\ast_0(2400)$ two-pole structure and the SU(3) pattern\footnote{Among other predictions, we point out that the lower $D^\ast_0(2400)$ resonance (located at $2105(8)-i102(12)\ \text{MeV}$) and the $D_{s0}^\ast(2317)$ state would be siblings forming a $\bar{3}$ SU(3) multiplet. The same can be said of the new $D^\ast_1$ resonance (located at $2240(6)-i 93(9)\ \text{MeV}$) and
the $D_{s1}(2460)$ state in the $1^+$ sector. All these features have counterparts in the bottom sector as well.}  of the $0^+$ and $1^+$ heavy-light sectors claimed in Ref.~\cite{Albaladejo:2016lbb} seem rather robust from the theoretical point of view. All these results reinforce the new paradigm to study the spectrum of heavy-light mesons~\cite{Du:2017zvv}, and that questions its traditional interpretation in terms of  constituent $Q\bar q$  degrees of freedom.

\subsection{\boldmath $DK$ S-wave Phase shifts}
\begin{figure*}[t]
\begin{tabular}{cc}
\includegraphics[width=0.45\textwidth,keepaspectratio]{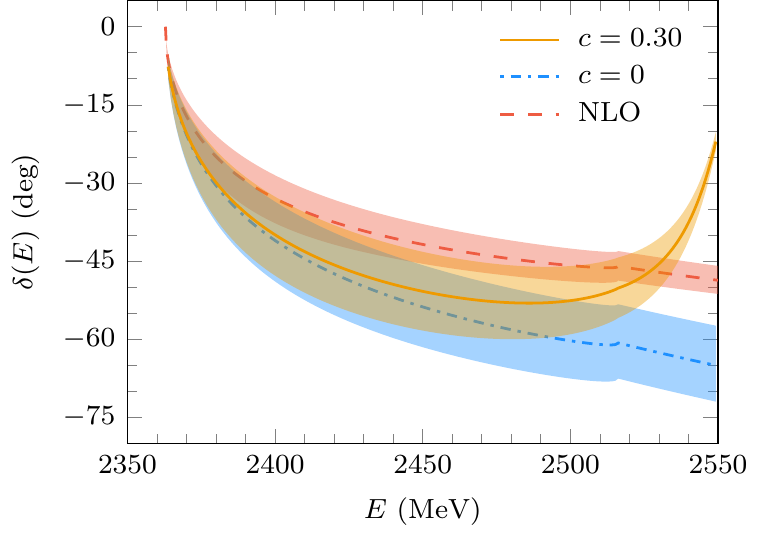} & \includegraphics[width=0.45\textwidth,keepaspectratio]{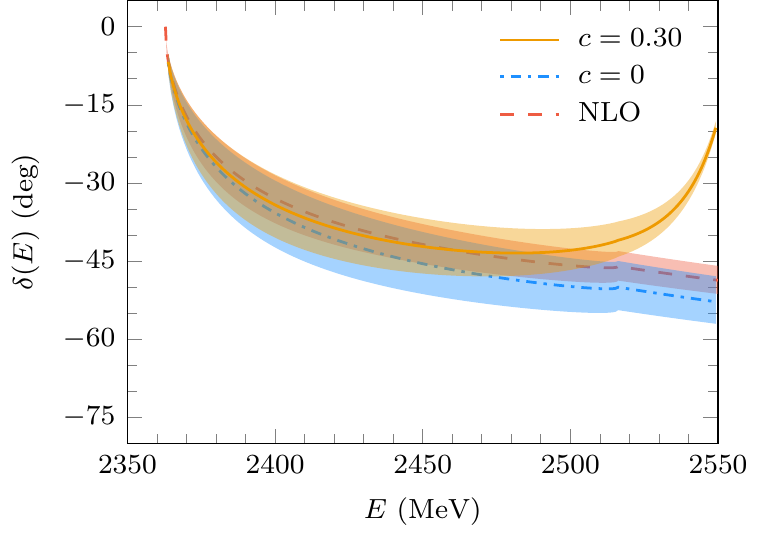}\\
\includegraphics[width=0.45\textwidth,keepaspectratio]{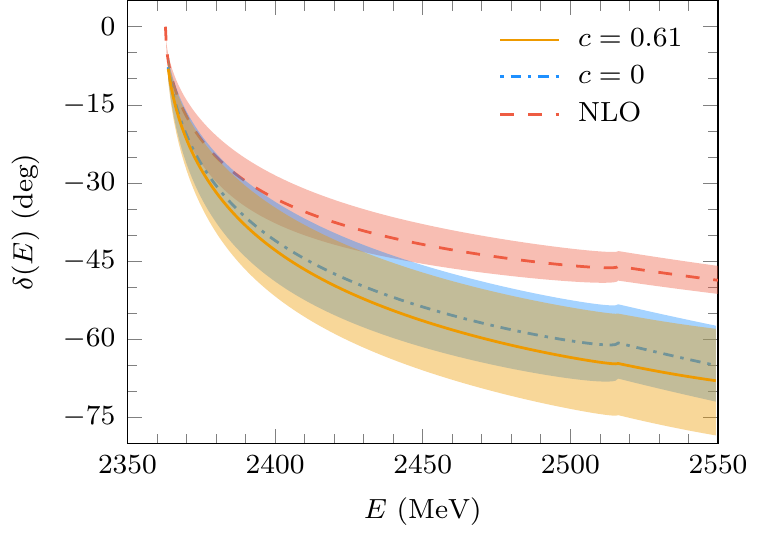} & \includegraphics[width=0.45\textwidth,keepaspectratio]{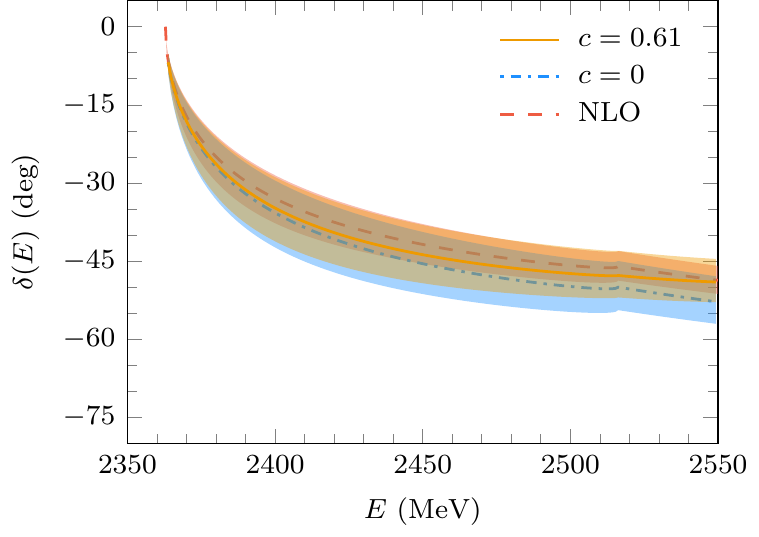}
\end{tabular}
\includegraphics[width=0.45\textwidth,keepaspectratio]{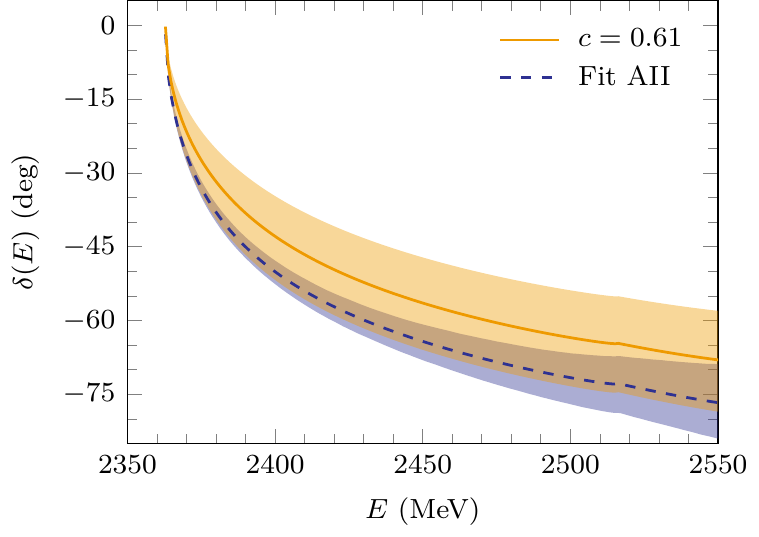}
\caption{First two rows of plots: energy dependence of the $DK$ S-wave phase shifts obtained using the chiral unitarized approach of Refs.~\cite{Liu:2012zya,Guo:2008gp} (NLO) compared with the results deduced including the exchange of a CQM state of mass 2511 MeV (Set A).  Predictions for three  different values of the dimensionless LEC $c$, that controls the
strength of the coupling of the bare CQM meson and the $P\phi-$pair, are depicted. Note that in the case $c=0$, the approach reduces to the unitarized LO HMChPT approach followed in \cite{Albaladejo:2016ztm} for the bottom-strange sector, except that in this latter case the $\bar B_s\eta$ channel was not considered. In the left (right) panels  the UV divergences, that appear in the unitarization of the LO HMChPT+CQM amplitudes, have been renormalized using a Gaussian cutoff (subtraction constant). For details  see Tables~\ref{table:nlo-comparison} and \ref{table:nlo-comparison-sub}. The UV behaviour of the NLO unitarized amplitudes is always renormalized  using subtraction constants~\cite{Liu:2012zya,Guo:2008gp}. Statistical 68\%-CL error-bands are generated from the uncertainties of the LECs that enter in each scheme.\label{fig:phase-shifts-comparison}
Bottom plot: comparison of phase-shifts for $c=0.61$ using two different UV Gaussian cutoffs, $0.71^{+0.07}_{-0.06}$ GeV and  $0.81^{+0.17}_{-0.12}$  GeV. The first value corresponds to Fit AII of Table~\ref{tab:Results}, and the uncertainty of $\pm 0.09$ quoted there for the the LEC $c$ is also  considered to obtain the statistical 68\%-CL error-band. The second UV cutoff is that associated to $c=0.61$ in 
Table~\ref{table:nlo-comparison}, and the deduced phase-shifts are also shown in the left panel of the second row of plots in this figure. Both sets of LECs produce different  $\Dvd$ masses, $2331\pm 3$ MeV (Table~\ref{tab:Results}) and $2315_{-28}^{+18}$ MeV (Table~\ref{table:nlo-comparison}), respectively. In all cases, physical $D$, $D_s$, $K$ and $\eta$ masses have been used.}  
\end{figure*}
\begin{table*}
\centering
\begin{ruledtabular}
\def\arraystretch{1.3}
\caption{Three different sets of $c-\Lambda$ pairs leading to the same $D^\ast_{s0}(2317)$ mass ($2315_{-28}^{+18}$) obtained from the unitarized LO HMChPT+CQM approach. The UV cutoff $\Lambda$  is fitted to reproduce the above mass, that it is deduced from the unitarized NLO HMChPT $T-$matrix~\cite{Liu:2012zya}. 
The uncertainties in the $\Dvd$ mass, inherited  from the errors on  the LECs, are taken into account and lead to the errors on $\Lambda$. We first generate synthetic sets of LECs, according to the correlated error distributions given in Ref.~\cite{Liu:2012zya}. Next, we find the position of the $D^\ast_{s0}(2317)$ pole for  each set of LECs, and finally, for each parameter $c$,  we tune the  UV cutoff to reproduce this mass. In this way, we determine the error distributions of the cutoffs that are also used to estimate the uncertainties on the derived quantities (S-wave $DK$ scattering length, mass and  width of the higher --dressed-- resonance, $\Dvd$ molecular probabilities  and couplings) given in the table. Predictions for the latter quantities from the unitarized NLO HMChPT approach~\cite{Guo:2008gp, Liu:2012zya} are compiled in the last row. Finally note that the choice $c=0.61$ is motivated by the Set A of results in Table~\ref{tab:Results}, which $0^+$ CQM bare mass is always employed in this table. 
\label{table:nlo-comparison}}
\begin{tabular}{cccccccccc}
$c$ & $\Lambda$ & $M_{D^\ast_{s0}}$ & $g_{DK}$ & $g_{D_s\eta}$ & $P_{DK}$ & $P_{D_s\eta}$ & $a_{DK}$ & $M_R$ & $\Gamma_R$\\
    & [GeV] & [MeV] & [GeV] & [GeV] & [\%] &  [\%] & [fm] & [MeV] & [MeV] \\
    \hline
$0$ & $1.19_{-0.13}^{+0.18}$ & $2315_{-28}^{+18}$ & $11.3_{-1.0}^{+0.7}$ & $6.7_{-0.4}^{+0.2}$ & $61_{-10}^{+8}$ & $5.9_{-0.7}^{+0.3}$& $-1.00_{-0.30}^{+0.23}$  & $-$ &  $-$ \\
$0.30$ & $1.05_{-0.13}^{+0.18}$ & $2315_{-28}^{+18}$ & $11.2_{-1.0}^{+0.7}$ & $6.3_{-0.4}^{+0.3}$ & $58_{-9}^{+8}$ & $4.7^{+0.4}_{-0.7}$ & $-0.98_{-0.30}^{+0.23}$ & $2557_{-2}^{+1}$ & $28_{-4}^{+2}$ \\
$0.61$ & $0.81_{-0.12}^{+0.17}$ & $2315_{-28}^{+18}$ & $11.7_{-1.0}^{+0.7}$ & $6.0_{-0.5}^{+0.3}$ & $57_{-9}^{+8}$ & $3.0_{-0.7}^{+0.6}$ & $-1.04_{-0.31}^{+0.24}$ & $2686_{-5}^{+7} $ & $90_{-6}^{+2}$ \\
\multicolumn{2}{c}{NLO}& $2315_{-28}^{+18}$ & $9.5^{+1.2}_{-1.1}$ & $7.5\pm 0.5$ & $54\pm 4$ & $13^{+6}_{-10}$ & $-0.84^{+0.17}_{-0.23}$ & $-$ & $-$\\
\end{tabular}
\end{ruledtabular}
\end{table*}
Here we will show predictions for $DK$ S-wave scattering phase-shifts [see Eq.~\eqref{eq:T-matrix-parametrization}], and will take the opportunity to discuss few aspects of the renormalization dependence of the scheme. For the sake of brevity, we will not address the rest of channels.  In this subsection, we will always show results obtained for infinite volume, using physical meson masses, and the Set A of bare masses to incorporate the CQM degrees of freedom.

The behavior of the phase shifts at threshold depends strongly on the position of the S-wave $DK$ pole, as can be seen in the bottom panel of Fig.~\ref{fig:phase-shifts-comparison}.  Thus, and to make the discussion meaningful, we will consider approaches leading to the same  $\Dvd$ mass value, $2315_{-28}^{+18}$, obtained in the unitarized NLO HMChPT approach~\cite{Guo:2008gp, Liu:2012zya}, and whose central value is quite close to the experimental one. Details can be found in Table~\ref{table:nlo-comparison}, while the related phase shifts are shown in the left panels of the first two rows in Fig.~\ref{fig:phase-shifts-comparison}.

Looking at the results of Table~\ref{table:nlo-comparison}, we first stress the dependence of the UV cutoff on the LEC $c$, or viceversa the dependence of $c$ on $\Lambda$, for a fixed CQM bare mass. The latter should be also understood as a renormalization scheme dependent quantity. All these three parameters ($c,\Lambda, \mbare$) should effectively incorporate higher order chiral corrections beyond LO, and not accounted by the unitarity loops (see also the discussion in Appendix~\ref{app:integrate-out}). One expects that these further corrections should be rather independent of the short distance physics at energies moderately far from threshold. However predictions might significantly differ, lets us say above 2450 or 2500 MeV. Indeed, we have an indication from the masses and widths of the possible second (higher resonance) state compiled in the table. We see that for $c=0.3$, a narrow resonance (28 MeV) is generated at around  2557 MeV,  that would produce clear signatures, not seen,  in the second 
energy level calculated in Ref.~\cite{Bali:2017pdv}. The other 
schemes either do not generate any resonance ($c=0$ and NLO) or it is located close to 2700 GeV ($c=0.61$) and it is sufficiently broad to become unnoticed for energies  below 2600 MeV.  The conclusion is that the $c=0.3$ predictions above 2500 MeV are unreliable, and that the exact location of the resonance generated for $c=0.61$ is not well constrained by the data of Ref.~\cite{Bali:2017pdv}. On the other hand,  we appreciate a variation pattern for $c$ consistent with the physical interpretation of this LEC, and thus the total molecular probability decreases from $(67\pm 10)\%$ down to $(60\pm 9) \%$, when $c$ varies from 0 to 0.61. Scattering lengths are mostly determined by the common $\Dvd$ mass and do not show statistically significant variations, as it also occurs  for the couplings. Paying attention now to the NLO results, we find that they coincide reasonably well with those found using LO HMChPT amplitudes, thanks to the freedom in the latter to  re-adjust the cutoff. Finally, it could  be 
surprising that the $DK$ molecular probability, obtained within the NLO scheme, is only $54\pm 4\%$, when a value of $\sim 70$\% was claimed  in the original work of Ref.~\cite{Liu:2012zya}. This discrepancy is due to  the use in the latter work of the Weinberg compositeness rule \cite{Weinberg:1965zz}, that provides the molecular probability in terms of the scattering length and  the $DK$ wave number $\gamma_B$,
\begin{equation}
 P_{DK} \sim -\frac{a_{DK}}{a_{DK}+2/\gamma_B}, \quad \gamma_B=\sqrt{2\mu B}, \label{eq:prob-wb} 
\end{equation}
which leads to one, in the limiting case when the scattering length is approximated by $-\gamma_B^{-1}$ (very loosely bound states), neglecting finite effective range corrections. Indeed, using the relation of Eq.~\eqref{eq:prob-wb}, one obtains $P_{DK}\sim 70\pm 4$ \%, as obtained in Ref.~\cite{Liu:2012zya}. Note, however, that  Eq.~\eqref{eq:prob-wb} has  corrections due to the $\Dvd$ binding energy~\cite{Torres:2014vna}, which is not too small ($\sim 46$ MeV), and possible inelastic effects~\cite{Baru:2003qq}, which should bring the $70\%$ down to the more accurate estimate found in the present work.

Coming back to the phase shifts shown in the left panels of Fig.~\ref{fig:phase-shifts-comparison}, we see that different schemes produce compatible phase shifts close to threshold, while the differences become larger as the energy increases. Thus for instance, the $c=0.3$ phase shifts suddenly change curvature above $2500\ \text{MeV}$,  but as we discussed above such behaviour, produced for a narrow resonance at $2557\ \text{MeV}$ (see Table~\ref{table:nlo-comparison}), is not compatible with the higher $0^+$ LQCD energy level reported in Ref.~\cite{Bali:2017pdv}. Above $2450\ \text{MeV}$ the rest of schemes lead to differences in the phase shifts of few degrees, and at most of around $15^\circ$  at $2550\ \text{MeV}$. However, taking into account the uncertainties of the different predictions, phase shifts are almost compatible up to this latter energy. 

\begin{table*}
\centering
\begin{ruledtabular}
\def\arraystretch{1.3}
\caption{Same as Table~\ref{table:nlo-comparison}, but using  a different  procedure to renormalize the meson loop matrix $G(s)$. Here, as in Refs.~\cite{Liu:2012zya,Guo:2008gp}, we have used dimensional regularization with a subtraction constant $\alpha$ (see Eq.~\eqref{eq:rs} and the related discussion), which is adjusted to reproduce the $\Dvd$ mass.\label{table:nlo-comparison-sub}}
\begin{tabular}{cccccccccc}
$c$ & $\alpha$ & $M_{D^\ast_{s0}}$ & $g_{DK}$ & $g_{D_s\eta}$ & $P_{DK}$ & $P_{D_s\eta}$ & $a_{DK}$ & $M_R$ & $\Gamma_R$\\
    &  & [MeV] & [GeV] & [GeV] & [\%] &  [\%] & [fm] & [MeV] & [MeV] \\
    \hline
$0$ & $-1.87_{-0.13}^{+0.10}$ & $2315_{-28}^{+18}$ & $10.2_{-0.9}^{+0.7}$ & $6.6_{-0.3}^{+0.2}$ & $63_{-8}^{+7}$ & $10_{-0.4}^{+0.1}$& $-0.90_{-0.27}^{+0.21}$  & $-$ &  $-$\\
$0.30$ & $-1.77_{-0.14}^{+0.11}$ & $2315_{-28}^{+18}$ & $10.0_{-0.9}^{+0.7}$ & $6.2_{-0.3}^{+0.2}$ & $60_{-7}^{+6}$ & $9.0_{-0.4}^{+0.1}$ & $-0.87_{-0.27}^{+0.20}$ & $2559.8_{-0.4}^{+0.2}$ & $23_{-3}^{+2}$ \\
$0.61$ & $-1.60_{-0.14}^{+0.11}$ & $2315_{-28}^{+18}$ & $10.1_{-0.9}^{+0.8}$ & $5.8_{-0.3}^{+0.2}$ & $61_{-7}^{+6}$ & $8.0_{-0.4}^{+0.1}$ & $-0.88_{-0.27}^{+0.20}$ & $2704_{-2}^{+3} $ & $88_{-7}^{+5}$ \\
\multicolumn{2}{c}{NLO}& $2315_{-28}^{+18}$ & $9.5^{+1.2}_{-1.1}$ & $7.5\pm 0.5$ & $54\pm 4$ & $13^{+6}_{-10}$ & $-0.84^{+0.17}_{-0.23}$ & $-$ & $-$\\
\end{tabular}
\end{ruledtabular}
\end{table*}
We now discuss about the dependence of the analysis on the regulator scheme. To this end, we consider the loop matrix function $G$ renormalized by  one subtraction, as in  Refs.~\cite{Guo:2008gp, Liu:2012zya}. Suppressing
the indices, the loop function is written for each channel as,
\begin{equation}
G(s)=\bar G(s) + G[(M+m)^2]. \label{eq:Gsubs}
\end{equation}
The finite function $\bar G(s)$ can be found in Eq.~(A9) of
Ref.~\cite{Nieves:2001wt}. On the other hand, the constant $G[(M+m)^2]$ contains the
UV logarithmic divergence. After renormalizing using  dimensional
regularization, one finds,
\begin{eqnarray}
G[(M+m)^2] &= & \frac{1}{16\pi^2} \Big ( a(\nu) +  \nonumber \\
&&  \frac{1}{M+m}  [ M \ln \frac{M^2}{\nu^2} + m \ln
\frac{m^2}{\nu^2} ]\Big) \label{eq:rs}
\end{eqnarray}
where $\nu$ is the scale of the dimensional regularization. Changes in
the scale are, in principle,  reabsorbed in the subtraction constant $a(\nu)$, so that
the results remain scale independent. Here we have taken $\nu= 1$ GeV and a common subtraction constant $a(1~\text{GeV}) = \alpha$ for both $DK$ and $D_s\eta$ channels, as in Refs.~\cite{Guo:2008gp, Liu:2012zya}. We 
have now constrained the LEC $\alpha$ to obtain the same $\Dvd$ mass ($2315_{-28}^{+18}$), as in Table~\ref{table:nlo-comparison}, for each value of the parameter $c$. The results 
are presented in Table~\ref{table:nlo-comparison-sub}. Comparing these latter results with those in Table~\ref{table:nlo-comparison}, we see that the predictions, within uncertainties, 
are consistent in the two renormalization schemes. There is a slight dependence, and the $\Dvd$ coupling to $DK$ and the modulus of the scattering length are smaller and closer to those deduced from the  NLO approach. Molecular probabilities are somewhat larger, specially $P_{D_s\eta}$ that becomes almost twice bigger. As a consequence, the total hadronic molecular component is now roughly $ 70 \pm 10$\%. The dependence on $c$ follows a similar pattern as in Table~\ref{table:nlo-comparison}, and the importance of the $D_s\eta$ channel decreases as the value of the LEC $c$ increases. Finally, the results concerning the higher--dressed--resonance pole position are similar in the two renormalization schemes. Thus from the previous discussion, the $c=0.3$ predictions for energies above $2500\ \text{MeV}$ turn out to be little reliable also in this scheme.

The  phase shifts deduced from the various possibilities  discussed in Table~\ref{table:nlo-comparison-sub} are shown in the right panels of the first two rows in Fig.~\ref{fig:phase-shifts-comparison}. We see, the $c=0.3$ phase shift above 2500 MeV presents the same pathologies as in the left top panel, where an UV Gaussian cutoff is used. It is interesting to note that the $c=0$ and $c=0.61$ phase shifts have smaller errors, the two sets of phase shifts are still statistically compatible, but in addition, they now agree quite well with the NLO predictions, and also qualitatively with those found in Ref.~\cite{Guo:2018kno}.  Thus, the renormalization scheme dependence, while it is not much relevant for the $\Dvd$ properties, turns out more important for the phase shift at energies above 2450 MeV, as one could reasonably expect from the previous discussions. This clearly could be regarded as a source of systematic uncertainty, though up to 2550 MeV remains smaller/comparable to the other uncertainties of the 
predictions accounted for the error bands depicted in Fig.~\ref{fig:phase-shifts-comparison}.


\section{Summary and concluding remarks}
\label{sec:concl}

In this work we have first carried out a coupled channel study of the $0^+$ and $1^+$ charm-strange meson sectors employing a chiral unitary approach based on LO HMChPT $P^{(\ast)}\phi$ interactions, 
and that incorporates, consistently with HQSS, the interplay between intermediate CQM  bare $c\bar s$ and $P^{(*)}\phi$ degrees of freedom. We have extended the scheme to finite volumes and fixed the strength of the coupling between both types of degrees of freedom to the available $0^+$ and $1^+$ LQCD energy-levels~\cite{Bali:2017pdv}, which we have successfully described.  On the other hand and at variance to the situation in the bottom-sector  reported in Ref.~\cite{Albaladejo:2016ztm}, we have found that the $0^+$ and $1^+$ CQM bare masses (denoted as Set B in this work) obtained in Ref.~\cite{Segovia:2012yh} using the one-loop corrections to the CQM OGE potential proposed in Ref.~\cite{Gupta:1994mw}, lead to a really  poor description of the LQCD data. This is because the HQSS breaking corrections induced by this modification of the OGE potential are inconsistent with the LQCD energy levels calculated in Ref.~\cite{Bali:2017pdv}.

We have  estimated the size of the  the $D^{(\ast)}K$ two-meson components in the $D^\ast_{s0}(2317)$ and $D_{s1}(2460)$, and conclude that these states have a predominantly hadronic-molecular structure. Furthermore,  we have observed a quite significant dependence of the lowest-lying LQCD energy-levels of Ref.~\cite{Bali:2017pdv} on the pion mass, which is difficult to accommodate by a dominant CQM $c\bar s$ component. This is, however,  consistent with having a large influence of the $P^{(*)}\phi$ loops in the $D_{s0}^\ast(2317)$ and $D_{s1}(2460)$ structure.

In addition, we have found one extra resonance, in both the $0^+$ and $1^+$ sectors, arising from the dressed CQM states. Our predictions for these states are not as robust as those for the low lying $D_{s0}^\ast(2317)$  and $D_{s1}(2460)$, and moreover they are relatively broad, which might complicate their discovery. Some experimental efforts are needed  to clarify their possible existence.

The LEC $c$ depends on the radial quantum number, but not on the heavy flavor, up to $\Lambda_{QCD}/m_Q$ corrections. Thus, the value determined here for this parameter should be similar to that found in  
the bottom-strange  sector in  Ref.~\cite{Albaladejo:2016ztm}. There, it was obtained $c=0.75(6)$, which is quite compatible with the values in the range $0.52$--$0.70$ found in this work for the Set A of CQM bare masses. Note that in addition to heavy flavor symmetry breaking corrections, there might be also some discretization errors. Nevertheless, we have  shown that taking into account explicitly the exchange of (bare) CQM mesons is not fundamental to describe the $D_{s0}^\ast(2317)$ and $D_{s1}(2460)$ states, since such contributions can be accommodated by appropriately modifying the finite contributions derived from short-distance physics. This is natural because the CQM bare states lie far from the latter physical states, for which the unitarity meson loops play a fundamental role.

We have discussed how the approach followed here, where the two meson loop function is computed in a finite volume, the unknown LECs are determined 
from fits to the LQCD data, and  finally  poles are searched for in the infinite volume unitarized amplitudes using physical meson masses, provides a theoretically sound tool to analyze the LQCD energy-levels. We have shown that such procedure leads to more precise predictions that those obtained via the L\"uscher's relation using the effective range approximation.

We have also calculated the energy-levels obtained from the unitarized HMChPT NLO amplitudes derived in Ref.~\cite{Guo:2008gp}, without including any contribution from the exchanges of CQM bare poles.  We have shown (Fig.~\ref{fig:levels-NLO}) that this scheme provides a more than acceptable description  of the $0^+$ and $1^+$ LQCD energy levels of Ref.~\cite{Bali:2017pdv}, despite having fixed all LECs to the values previously determined in Ref.~\cite{Liu:2012zya} (not fitted to the energy levels). These findings, together with the  similar good description found in Ref.~\cite{Albaladejo:2016lbb} of the $(S,I)=(0,1/2)$ LQCD low-lying energy levels calculated in Ref.~\cite{Moir:2016srx}, provide a great support for the amplitudes obtained after unitarizing the NLO HMChPT amplitudes derived in Refs.~\cite{Guo:2008gp,Liu:2012zya}. Hence, the $D^\ast_0(2400)$ two-pole structure and the SU(3) pattern of the $0^+$ and $1^+$ heavy-light sectors claimed in Ref.~\cite{Albaladejo:2016lbb} seem rather robust from the 
theoretical point of view. All these results reinforce a  
new paradigm to study the spectrum of heavy-light mesons~\cite{Du:2017zvv}, that questions its traditional interpretation in terms of  constituent $Q\bar q$  degrees of freedom.

Finally, we have predicted S-wave $DK$ phase shifts and  discussed few aspects of the renormalization dependence of our results.

\appendix

\section{HMChPT NLO LECs and bare CQM  pole exchange potential}
\label{app:integrate-out}
The contribution of resonance exchanges to the  LECs in the NLO and NNLO
chiral Lagrangian for the $P\phi$ interaction was discussed at length in Ref.~\cite{Du:2016tgp}. There, it was shown that these LECs receive contributions from
exchanging the scalar charmed mesons, the light-flavor vector, scalar, and tensor mesons. The exchanged charm mesons are those related to the physical resonances. Here, however, the bare CQM states do not correspond to the physical states, which, as we have seen, are mostly generated by unitarity loops. Thus, it is not evident how to proceed in our context along  the lines discussed in Ref.~\cite{Du:2016tgp}.

However, one could still relate the LECs appearing at NLO in the HMChPT amplitudes derived in Refs.~\cite{Guo:2008gp,Liu:2012zya} to the bare mass of the CQM $c\bar{s}$ state and its coupling  to the $P^{(*)}\phi$ degrees of freedom.  To make  the numerical comparison meaningful, one should take into account that full elastic unitarity is restored, through Eq.~(\ref{eq:InvTMatrix}), where $G(s)$  is the renormalized
two-meson loop function. For simplicity, we will consider the single channel $DK$ in the chiral limit, and  the LO+CQM approach with the loop  function $G$ renormalized by  one subtraction, as in  Refs.~\cite{Guo:2008gp, Liu:2012zya}, and discussed here in Eqs.~(\ref{eq:Gsubs}) and (\ref{eq:rs}). We define an effective potential 
$V_{\rm eff}(s) = \left(V^{-1}_{\rm LO+CQM}(s)+\alpha_c\right)^{-1}$, which is given in terms of the HMChPT LO+ bare CQM pole exchange potentials and  the additional constant $\alpha_c$. The latter  is defined as the  difference between the value found for $G[(M+m)^2]$ in \cite{Liu:2012zya} and that determined here using  $\alpha=-1.60_{-0.14}^{+0.11}$, subtraction constant\footnote{There is here a subtlety, and  in practice we use $\alpha=-1.42^{+0.13}_{-0.16}$. This is because we need to use this modified value of $\alpha$ to get the same value for the  $D_{s0}^\ast(2317)$ mass ($2336^{+14}_{-21}$ MeV) as that obtained with the  LECs of  Ref.~\cite{Liu:2012zya} when the $D_s\eta$ channel is switched off. Note that  in Table~\ref{table:nlo-comparison-sub}, $D_s\eta$ coupled channel effects are taken into account. The behavior of the $DK$ amplitude close to threshold depends significantly on the mass of the pole and therefore for numerical purposes this re-adjustment is needed.  } corresponding to
$c = 0.61$ in Table~\ref{table:nlo-comparison-sub}.  We make an expansion in the kaon energy and after matching the quadratic coefficients of $V_{\rm eff}$  and of the irreducible amplitude obtained in \cite{Liu:2012zya}, we readily find
\begin{equation}
 \frac{c^2 \mbare /M}{1-\mbare^2/M^2} - \alpha_c\frac{M^2}{f^2} \sim -\left ( h_2-h_3 \right)  - \left( h_4-2h_5 \right) M^2,
\end{equation}
where $h_i$ are LECs introduced in Refs.~\cite{Guo:2008gp,Liu:2012zya}. Left and right hand sides of the above equation give $0.57^{+0.29}_{-0.27}
$ and $0.52^{+0.22}_{-0.21}$, respectively, showing a remarkable agreement. 

%
\begin{acknowledgments}
We warmly thank F.-K. Guo, D. R. Entem and  E. Oset  for enlightening discussions, and in addition we are also quite grateful to  F.-K. Guo, who  gave us the NLO LECs distributions used in this work. 
P.~F.-S. acknowledges financial  support from the 
"Ayudas para contratos predoctorales para la formaci\'on de doctores" 
program (BES-2015-072049) from the Spanish MINECO and ESF. 
P.~G.~O. acknowledges financial support from the Spanish MINECO's "Juan de la Cierva-Incorporaci\'on" programme with grant agreement no.~IJCI-2016-28525 and from Junta de Castilla 
y Le\'on and European ERDF funds under Contract no.~SA041U16.
This work is supported by the Spanish MINECO and European ERDF funds under the contracts FIS2014-51948-C2-1-P, FIS2017-84038-
C2-1-P and SEV-2014-0398.
\end{acknowledgments}
\bibliography{Ds0-HQSS}
\end{document}